\documentclass{article}

\usepackage[english]{babel}

\usepackage[letterpaper,top=2cm,bottom=2cm,left=3cm,right=3cm,marginparwidth=1.75cm]{geometry}
\usepackage{setspace}

\usepackage{amsmath}
\usepackage{amssymb}
\usepackage{graphicx}
\usepackage{dsfont}
\usepackage{mathdots}
\usepackage[colorlinks=true, allcolors=blue]{hyperref}
\usepackage{xcolor}

\usepackage{array}
\usepackage{multirow}

\usepackage{standalone}

\usepackage{tikz}
\usetikzlibrary{positioning}
\usetikzlibrary{arrows, fit}
\usetikzlibrary {shapes.geometric}

\newcommand{\U}{\text{U}}
\newcommand{\SU}{\text{SU}}
\newcommand{\USp}{\text{USp}}
\newcommand{\SO}{\text{SO}}

\newcommand{\SL}{\text{SL}}
\newcommand{\PSL}{\text{PSL}}
\newcommand{\Dic}{\text{Dic}}
\newcommand{\Tr}{\text{Tr}}

\newcommand{\beq}{\begin{equation}}
\newcommand{\eeq}{\end{equation}}

\newcommand\addvmargin[1]{
  \node[fit=(current bounding box),inner ysep=#1,inner xsep=0]{};
}

\newcommand\blfootnote[1]{%
  \begingroup
  \renewcommand\thefootnote{}\footnote{\hspace{-6mm}#1}%
  \addtocounter{footnote}{-1}%
  \endgroup
}

\numberwithin{equation}{section}

\title{Origami with a Twist: Twisted Holography of Four-Dimensional $\mathcal{N}=2$ Orientifold Theories}
\author{Jacob Abajian}

\begin{document}
\begin{center}
$$$$
{\huge
Origami with a Twist: Twisted Holography \\
of Four-Dimensional $\mathcal{N}=2$ Orientifold Theories
\par}
\vspace{1.2cm}

{\Large Jacob Abajian}
\blfootnote{\tt{jacobmabajian@gmail.com}}
\\ \vspace{0.5cm}
{\it Perimeter Institute for Theoretical Physics,
Waterloo, Ontario N2L 2Y5, Canada}\\
{\it Department of Physics \& Astronomy, University of Waterloo, Waterloo, Ontario N2L 3G1, Canada}
\end{center}
\vspace{2cm}

\begin{abstract}
We consider $\mathcal{N}=2$ superconformal gauge theories in four dimensions. We explain how these quiver gauge theories arise as low-energy worldvolume theories of D3-branes on orientifolds. Then, we examine their associated chiral algebras, and propose novel examples of twisted holographic dualities arising in the large-N limit. These dualities involve topological strings in the bulk, which is a Calabi-Yau threefold taking the form $\SL_2 \mathbb{C} / \Gamma$.
\end{abstract}

\newpage

\section{Introduction}

In recent years, the concept of symmetry has been expanded in various ways, and it has become apparent that the key ingredient is the existence of topological operators \cite{Gaiotto:2014kfa}.
Most of the familiar consequences of symmetry, such as Ward identities and superselection rules, follow simply from the existence of such operators (even if the operators do not possess other properties common to traditional symmetries, e.g. invertibility or codimensionality-1).

In \cite{Beem:2013sza}, the authors introduced a chiral algebra associated to each four-dimensional $\mathcal{N}=2$ superconformal field theory (SCFT).
This chiral algebra consists of a subset of the local operators of the four-dimensional theory, which are restricted to lie within a two-dimensional plane.
The operators then only depend on their position within that plane holomorphically, $\mathcal{O}(z, \bar{z}; w = \bar{w} = 0) = \mathcal{O}(z)$, so we can form line operators of the form $\int dz f(z) \mathcal{O}(z)$ for any holomorphic function $f$ which are topological within the two-dimensional plane.
We see then that the chiral algebra is a novel manifestation of symmetry, in the general sense described above.

The chiral algebra arises from a ``twist'' of the SCFT by $\mathcal{Q}+\mathcal{S}$, where $\mathcal{Q}$ and $\mathcal{S}$ are a supersymmetry and superconformal generator respectively.
The operator $\mathcal{Q}+\mathcal{S}$ obeys $(\mathcal{Q}+\mathcal{S})^2 = 0$, and it acts as a derivation on the SCFT's space of local operators.
To ``twist'' the theory by $\mathcal{Q}+\mathcal{S}$ means to pass to cohomology with respect to $\mathcal{Q}+\mathcal{S}$.
This twist isolates a protected subsector of the theory, whose OPE's are invariant under deformations of the theory that preserve $\mathcal{N}=2$ superconformal symmetry, and which obeys the structure of a chiral algebra (i.e. a set of operators that depend on two-dimenisonal position $z$ holomorphically, with an OPE that obeys the crossing equation).

In light of the AdS/CFT correspondence, it is natural to ask what the holographic interpretation of the chiral algebra associated to a holographic SCFT is.
Large-N $\mathcal{N}=2$ superconformal gauge theories with an 't Hooft expansion have dual descriptions as type IIB superstring theory in asymptotically AdS$_5$ backgrounds.
An analog of ``twisting'' was introduced for supergravity theories in \cite{Costello:2016mgj}.
The idea is that twisting supergravity corresponds to turning on a background for a bosonic ghost field.
At least at a perturbative level, this isolates a subsector of the supergravity theory in which a certain amount of supersymmetry is preserved. This is because only for such supersymmetric configurations is it possible to have a nonzero bosonic ghost field while obeying the equations of motion.

It was also proposed that this notion of ``spacetime twisting'' is related, in the case of the superstring, to that of worldsheet twisting.
That is, the twist of a supergravity theory arising as a low-energy limit of superstring theory should be related to the physics of a topological string, which arises as the worldsheet twist of the physical superstring.
Of particular interest for the present story is the type IIB superstring, which admits a worldsheet twist that is described by the B-model topological string.
The spacetime physics of B-model topological strings is the Kodaira-Spencer theory of gravity a.k.a. BCOV theory \cite{Bershadsky:1993cx, Costello:2012cy, Costello:2015xsa}.

The twist by $\mathcal{Q}+\mathcal{S}$ of an $\mathcal{N}=2$ SCFT is equivalent to an $\Omega$-deformation of Kapustin's holomorphic-topological twist \cite{Kapustin:2006hi, Oh:2019bgz, Jeong:2019pzg}.
For a holographic SCFT, this corresponds to introducing an $\Omega$-background in the bulk supergravity theory, which localizes the theory onto a three complex-dimensional subspace of the bulk.
So, we expect that the holographic dual of the $\mathcal{Q}+\mathcal{S}$ twist of a large-N $\mathcal{N}=2$ superconformal gauge theory is the BCOV theory in three complex dimensions.
This proposal was made precise in \cite{Costello:2018zrm} (see also \cite{Bonetti:2016nma}), wherein it was argued that the chiral algebra arising as the $\mathcal{Q}+\mathcal{S}$ twist of $\mathcal{N}=4$ super Yang-Mills theory with gauge group $\SU(N)$ has a holographic description as BCOV theory on $\SL_2 \mathbb{C}$.

Four-dimensional $\mathcal{N}=4$ super Yang-Mills theory is the low-energy description of a stack of D3-branes in flat space.
The twist by $\mathcal{Q} + \mathcal{S}$ converts the bulk type IIB superstring theory on $\mathbb{R}^{10}$ into the B-model topological string on $\mathbb{C}^3$, and replaces the physical D3-branes with B-model D1-branes wrapping $\mathbb{C} \subset \mathbb{C}^3$.
The backreaction of these branes deforms $\mathbb{C}^3$ into $\SL_2 \mathbb{C}$.

It is natural to extend this proposal to the $\mathcal{N}=2$ superconformal quiver gauge theories that arise by placing D3-branes at the fixed locus of an orientifold.
In fact, the case of D3-branes probing the singularity of an \emph{orbifold} was described already in \cite{Costello:2018zrm}.
It is the main purpose of the present paper to extend the twisted holographic almanac to include those $\mathcal{N}=2$ SCFT's that come from D3-branes probing orientifolds with nontrivial orientation-reversing elements.

The behavior of branes probing orbifold and orientifold singularities has been a subject of great interest for a long time \cite{Douglas:1996sw, Landsteiner:1997vd, Kapustin:1998fa, Kakushadze:1998tr, Uranga:1998uj, Park:1998zh, Uranga:1999mb}.
Constructions of supersymmetric field theories in terms of these brane configurations have led to the solution of many such theories, in the form of the construction of their Seiberg-Witten curves of supersymmetric vacua \cite{Hanany:1996ie, Witten:1997sc, Katz:1997eq, Kapustin:1998xn} and understanding of their moduli spaces of couplings \cite{Gaiotto:2009we}.

The orientifolds that preserve $\mathcal{N}=2$ supersymmetry come in two distinct categories, depending on whether the orientifold produces an O3-plane or an O7-plane.

\subsection*{Orientifolds with an O7-plane}

We are interested in orientifolds that preserve four-dimensional $\mathcal{N}=2$ supersymmetry.
Among the symmetries of type IIB superstring theory by which we can quotient while preserving that amount of supersymmetry is $\Omega' \equiv \Omega R_{45} (-1)^{F_L}$, which acts by reversing worldsheet parity, reversing the sign of two coordinates (the ``4 and 5'' directions), and multiplication by a sign depending on the left-moving fermion number.
When $\Omega' \in \Gamma$, it is necessarily of the form $\Gamma = G_1 \oplus \Omega' G_1$, where $G_1 \subset \SU(2)$ is a discrete subgroup of the $\SU(2)$ acting on the 6, 7, 8, and 9 directions.
Orientifold groups of this form result in orientifolds contains an O7-plane, the fixed locus of $\Omega'$.
It is then necessary to include a stack of D7-branes to cancel the RR-charge of the O7-plane.

We find that upon twisting, the theory is localized entirely onto the O7-plane, such that the local physics is described by \emph{unoriented} topological strings \cite{Costello:2019jsy}, and the relevant vacuum solution is $\SL_2 \mathbb{C}/ G_1$.

\subsection*{Orientifolds with an O3-plane}

Another type of orientifold which also preserves four-dimensional $\mathcal{N}=2$ supersymmetry contains an O3-plane.
These orientifolds are determined by a choice of discrete subgroup $\Gamma \subset \SU(2)$, \emph{together} with a choice of index-2 subgroup $G_1 \subset \Gamma$.
This is the subgroup of orientation-preserving elements.
The complement $\Gamma \setminus G_1$ are orientation-reversing elements of the form $\Omega' \alpha$, where $\alpha \in \SU(2)$ and $\alpha^2 \in G_1$.

Upon twisting, we find that the bulk physics is described locally by oriented topological strings on $\SL_2 \mathbb{C} / \Gamma$.
The effect of including orientation-reversing elements in the orientifold group is to introduce twisted boundary conditions for the some of the bulk fields.

\vspace{5mm}

\noindent The rest of this note proceeds as follows: In section \ref{QuiverConstructions} we describe how the $\mathcal{N}=2$ superconformal field theories which admit a weakly coupled limit arise as low energy worldvolume theories for stacks of D3-branes probing various orientifolds.
Many of the constructions we describe have been considered previously in the literature (see e.g. \cite{Sen:1996vd, Douglas:1996js, Argyres:2002xc, Ennes:2000fu, Beccaria:2020hgy}), and our goal has been to organize the relevant possibilities into a complete list.
In section \ref{TwistedBackgrounds}, we describe the twisted holographic backgrounds which arise from the orientifolds, resolving some questions about which information of the physical holographic duality survives the twisting procedure.
We then conclude with a summary and some comments regarding possible future directions.

\section{Orientifold Constructions of $\mathcal{N}=2$ Gauge Theories}\label{QuiverConstructions}

A large class of four-dimensional $\mathcal{N}=2$ supersymmetric theories are those which admit weakly-coupled limits with Lagrangian descriptions in terms of gauge fields.
These theories were enumerated in \cite{Bhardwaj:2013qia}, and they have field content summarized by a quiver diagram, with each node corresponding to an $\mathcal{N}=2$ vectormultiplet and edges corresponding to hypermultiplets.
Since we are interested in theories which admit a holographic description in terms of type IIB superstrings, we restrict our attention to those theories that:
\begin{itemize}
\item Contain operators whose correlation functions admit an 't Hooft expansion (i.e. fields with two indices corresponding to gauge groups whose rank(s) can be taken to be large).
\item Have a global symmetry group that admits a finite rank limit as the ranks of the gauge groups are taken to be large.
\end{itemize}

These two requirements narrow our focus to families of finite-sized quivers, with nodes whose rank can be taken to be large, and which require only a finite number of hypermultiplets as the nodes are taken to be large (e.g., we exclude $\mathcal{N}=2$ super-QCD, since it has a number of fundamental flavor hypermultiplets $N_f$ that scales with the rank of the gauge group).

The $\mathcal{N}=2$ superconformal gauge theories we are interested in arise in various ways in string theory (with different constructions often related by dualities).
However, for the purpose of finding the holographic dual description, it is most useful to note that we can construct the $\mathcal{N}=2$ superconformal gauge theories by placing a stack of D3-branes at the fixed locus of an orientifold of flat space and considering the low-energy worldvolume theory.
The general form of an orientifold preserving four-dimensional $\mathcal{N}=2$ supersymmetry can be understood as follows:

The orientifold group $\Gamma$ can be split into an orientation-preserving index-two subgroup $G_1$, and its complement, a set of orientation-reversing elements $G_2$.
If we view flat ten-dimensional spacetime as $\mathbb{R}^4 \times \mathbb{R}^2 \times \mathbb{C}^2$, with D3-branes wrapping $\mathbb{R}^4$, then the orientation-preserving subgroup $G_1$ is a finite subgroup of the $\SU(2)$ acting on $\mathbb{C}^2$.
The finite subgroups of $\SU(2)$ fall into an ADE classification.
There two infinite families consisting of the cyclic groups $\mathbb{Z}_k$, and dicyclic groups $\Dic_k$, along with three exceptional groups: the binary tetrahedral group, binary octahedral group, and binary icosahedral group.

The orientation-reversing elements $g \in G_2$, of the orientifold group are of the form: 

\beq
g = \alpha \Omega'
\eeq
with $\alpha \in SU(2)$ acting on $\mathbb{C}^2$ and
\beq
\Omega' = \Omega R_{45} (-1)^{F_L}
\eeq
where $\Omega$ is the worldsheet parity operator, $R_{45}$ reverses the sign of the coordinates on the $\mathbb{R}^2$ ``4 and 5 directions'', and $F_L$ is the left-moving fermion number on the worldsheet.
If the orientifold group contains $\Omega'$, the orientifold group will be $\mathbb{Z}_2^{\Omega'} \times G_1$.
Since the fixed locus of $\Omega'$ is the origin in the $\mathbb{R}^2$ $4$ and $5$ directions, we see that the orientifold by $\mathbb{Z}_2^{\Omega'} \times G_1$ contains an O7-plane.
As O7-planes carry RR charge, we will need to include 8 (half-)D7-branes to cancel it and avoid an anomaly that would render the worldvolume theory non-conformal.
This case is of special interest, since it will be the one relevant for unoriented topological strings \cite{Costello:2019jsy}.

When considering a stack of $N$ D3-branes at the fixed point of an orientifold by $\mathbb{Z}_2^{\Omega'} \times G_1$, the low-energy theory consists of two sectors, coming from the massless modes of 3-3 and 3-7 open strings respectively.
Before the orientifold, we have $\mathcal{N}=4$ super Yang-Mills theory with gauge group $\U(N)$, and the 3-3 sector of the orientifolded theory is a truncation of $\mathcal{N}=4$ SYM obtained by imposing invariance under the orientifold group.

Before we can determine this projection, we need to define the action of the orientifold group on the fields, which entails a choice of projective unitary representation $\gamma : \Gamma \rightarrow U(N)$ for how the orientifold group acts on the Chan-Paton factors.
Then, imposing invariance under the orientation-preserving subgroup $G_1$ corresponds to the conditions:

\beq
A_{\mu} = \gamma_{g} A_{\mu}\gamma_{g}^\dag
\eeq

\beq
X^i = {R^i}_j(g) \gamma_{g}X^j \gamma_{g}^\dag
\eeq
for all $g \in G_1$, where $R$ is the fundamental representation of $SU(2)$.

The orientation-reversing elements of $\Gamma$ act on the fields through transposition in addition to conjugation by unitary matrices.
Invariance under $G_2$ imposes the conditions:

\beq
A_{\mu} = -\gamma_{\alpha \Omega'}(A_{\mu})^{\intercal}\gamma_{\alpha \Omega'}^\dag
\eeq

\beq
X^i = {R^i}_j(\alpha) \gamma_{\alpha \Omega'}(X^j)^{\intercal}\gamma_{\alpha \Omega'}^\dag
\eeq
where the extra minus sign in the vectormultiplet condition comes from the fact that the vectormultiplet's oscillator state is odd under worldsheet parity.
Closure of the orientifold group implies that if $g_1 \Omega', g_2 \Omega' \in G_2$, then $g_1 g_2 \in G_1$, and the representation $\gamma$ must satisfy

\beq
\gamma_{g_1 \Omega'} \left(\gamma_{g_2 \Omega'}^{\dag}\right)^{\intercal} = e^{i \varphi(g_1 \Omega',g_2 \Omega')} \gamma_{g_1 g_2}.
\eeq

\subsection{Review of Orbifold Constructions}

\begin{figure}
\begin{center}
\begin{tikzpicture}[->,>=stealth',shorten >=1pt,auto,node distance=3cm,
                    thick,main node/.style={circle,draw,font=\sffamily\Large\bfseries}]

  \node[main node] (1) {$m_0$};
  \node[main node] (2) [below right of=1] {$m_1$};
  \node[main node] (3) [below left of=2] {$m_2$};
  \node[main node] (4) [below left of=1] {$m_3$};
  
  \path[every node/.style={font=\sffamily\small}]
    (1) edge [bend left] node [left] {} (4)
        edge [bend left] node[left] {} (2)
    (2) edge [bend left] node [right] {} (1)
        edge [bend left] node [right] {} (3)
    (3) edge [bend left] node [right] {} (2)
        edge [bend left] node [right] {} (4)
    (4) edge [bend left] node [left] {} (3)
        edge [bend left] node[left] {} (1);
\end{tikzpicture}
\end{center}
\caption{The quiver diagram representing the field content of the $\mathbb{Z}_4$ orbifold. Nodes represent $\U(m_i)$ $\mathcal{N}=2$ vectormultiplets, and edges represent $\mathcal{N}=2$ hypermultiplets transforming in the $\textbf{fund} \otimes \overline{\textbf{fund}}$ representation of the two $\U(m_i)$ gauge groups connected by that edge.}
\label{Z4OrbQuiver}
\end{figure}
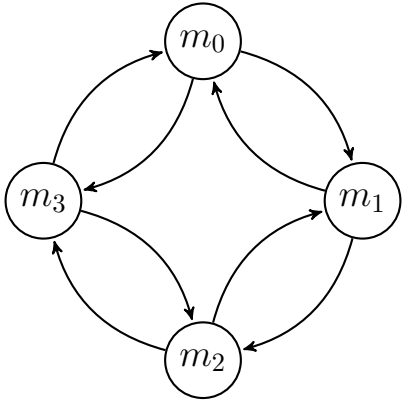

It has long been understood that the low-energy description of a stack of D-branes probing certain orbifold singularities is given by a quiver gauge theory \cite{Douglas:1996sw}.
In particular, four-dimensional $\mathcal{N}=2$ theories with semi-simple gauge groups of the form $\U(N_1) \times \dots \times \U(N_k)$, with hypermultiplets transforming in bifundamental representations arise by considering stacks of D3-branes on $\mathbb{R}^6 \times (\mathbb{C}^2/\Gamma)$.

In order to maintain $\mathcal{N}=2$ supersymmetry, we must have $\Gamma \subset \SU(2)$ acting on the $\mathbb{C}^2$ transverse to the D3-branes.
The condition that the states be invariant under the action of the orbifold group $\Gamma$ reduces the gauge group from $\U(\sum_{i} N_i)$ to $\U(N_1) \times \dots \times \U(N_k)$, and eliminates most of the hypermultiplets, keeping only those transforming under the bifundamental representations of neighboring $\SU(N_i)$ factors, as illustrated by a quiver diagram \cite{Douglas:1996sw}.

The simplest examples are the cyclic groups $\Gamma = \mathbb{Z}_k$. In this case, we obtain a theory described by a necklace quiver, see Figure \ref{Z4OrbQuiver}. This is also the only case in which the quiver diagram includes a loop. 
The holographic dual of this theory is type IIB superstring theory on AdS$_5\times(S^5/\mathbb{Z}_k)$, as described in \cite{Kachru:1998ys}.

In general, the orbifold theories are described by quivers in the shape of affine A, D, and E type Dynkin diagrams, which correspond to the (cyclic, dicyclic, and exceptional) finite subgroups of $\SU(2)$, see Table \ref{OrbifoldTable}.
In the 't Hooft limit, they are described holographically by type IIB superstrings on AdS$_5\times(S^5/\Gamma)$.

\subsection{$\mathbb{Z}_2^{\Omega'} \times \mathbb{Z}_k$}

Now consider the case $\Gamma = \mathbb{Z}_2^{\Omega'} \times \mathbb{Z}_k$, where the orientation-preserving subgroup is $G_1 = \mathbb{Z}_k \subset SU(2)$.
We will place $N$ D3-branes and 8 (half-)D7-branes at the fixed point of this orientifold.
Consider the 3-3 subsector of this theory, which arises as a truncation of four-dimensional $\U(N)$ $\mathcal{N}=4$ super Yang-Mills theory:

Denote the generator of $\mathbb{Z}_k$ by $\alpha$.
We need to specify how the orientifold group acts on the Chan-Paton factors of the open strings.
Specifically, we will have unitary matrices, $\gamma_{\Omega'}$ and $\gamma_{\alpha}$, such that:

\beq
\gamma_{\Omega'} (\gamma_{\Omega'})^* = \pm \mathds{1}; \hspace{1cm} (\gamma_{\alpha})^k = \mathds{1}.
\eeq

Up to a change of basis, we can always represent $\alpha$ as:

\beq
\gamma_{\alpha} = \left(\begin{array}{ccccc}
    \mathds{1}_{m_0} & 0 & 0 & \cdots & 0\\
    0 & \omega_{k} \mathds{1}_{m_1} & 0 & \cdots & 0 \\
    0 & 0 & \omega_{k}^2 \mathds{1}_{m_2} & \cdots & 0 \\
    \vdots & \vdots & \vdots & \ddots & \vdots \\
    0 & 0 & 0 & \cdots & \omega_k^{k-1} \mathds{1}_{m_{k-1}}
\end{array} \right)
\eeq
where $\omega_k = e^{2 \pi i/k}$. As is well known \cite{Douglas:1996sw}, invariance under the orbifold group $G_1 = \left<\alpha \right>$ projects the $U(N)$ vector multiplet onto the block-diagonal, resulting in a $U(m_0)\times U(m_1) \times \dots \times U(m_{k-1})$ gauge group:

\beq
A_{\mu} = \left(\begin{array}{ccccc}
    A_\mu^{(m_0)} & 0 & 0 & \cdots & 0\\
    0 & A_\mu^{(m_1)} & 0 & \cdots & 0 \\
    0 & 0 & A_\mu^{(m_2)} & \cdots & 0 \\
    \vdots & \vdots & \vdots & \ddots & \vdots \\
    0 & 0 & 0 & \cdots & A_\mu^{(m_{k-1})}
\end{array} \right)
\eeq
Since the hypermultiplet scalars carry an index in the $\mathbb{C}^2$ direction, they are projected into the bifundamental representation of ``neighboring'' $U(m_i)$ factors.
The field content after the $G_1$ orbifold can be nicely summarized in a quiver diagram, see Figure \ref{Z4OrbQuiver}.
We use \begin{tikzpicture}[thick,main node/.style={circle,draw,font=\sffamily\Large\bfseries}]
    \node[main node] (1) {$m$};
\end{tikzpicture} to represent $\U(m)$ gauge group factors, and arrows between nodes to represent the hypermultiplets transforming in the $\textbf{fund}\otimes \overline{\textbf{fund}}$ representation of the corresponding gauge group factors. 

Now we would like to perform a further orientifold by $\Omega'$.
At this point it becomes important whether $k$ is even or odd.

\subsubsection*{$k$ Even}

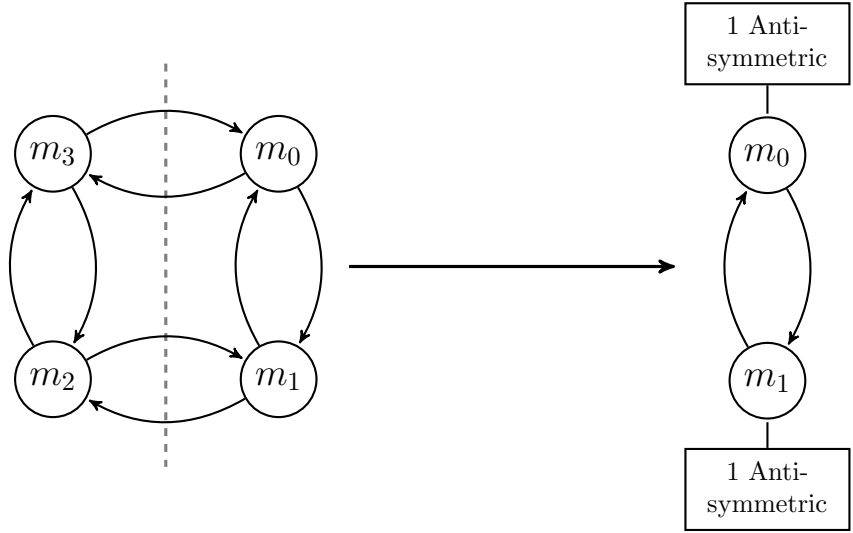
\begin{figure}
\begin{center}
\begin{tikzpicture}
\node[] (a) at (0,0) {
\begin{tikzpicture}[->,>=stealth',shorten >=1pt,auto,node distance=3cm,
                    thick,main node/.style={circle,draw,font=\sffamily\Large\bfseries}]

  \node[main node] (1) {$m_3$};
  \node[main node] (2) [right of=1] {$m_0$};
  \node[main node] (3) [below of=2] {$m_1$};
  \node[main node] (4) [below of=1] {$m_2$};

  \draw[dashed, -, very thick, black!50] (1.5,1.2) -- (1.5,-4.2);
  
  \path[every node/.style={font=\sffamily\small}]
    (1) edge [bend left] node [left] {} (4)
        edge [bend left] node[left] {} (2)
    (2) edge [bend left] node [right] {} (1)
        edge [bend left] node [right] {} (3)
    (3) edge [bend left] node [right] {} (2)
        edge [bend left] node [right] {} (4)
    (4) edge [bend left] node [left] {} (3)
        edge [bend left] node[left] {} (1);
        
\end{tikzpicture}
};
\node[] (b) at (8,0) {
\begin{tikzpicture}[->,>=stealth',shorten >=1pt,auto,node distance=3cm,
                    thick,main node/.style={circle,draw,font=\sffamily\Large\bfseries},square node/.style={rectangle,draw}]

  \node[square node] (1) {$\begin{array}{c} \text{1 Anti-} \\
  \text{symmetric}
  \end{array}$};
  \node[main node] (2) [below of=1,above=1cm] {$m_0$};
  \node[main node] (3) [below of=2] {$m_1$};
  \node[square node] (4) [below of=3,above=1cm] {$\begin{array}{c} \text{1 Anti-} \\
  \text{symmetric}
  \end{array}$};

  \path[every node/.style={font=\sffamily\small}]
    (1) edge[-] [left] node [left] {} (2)
    (2) edge [bend left] node [right] {} (3)
    (3) edge [bend left] node [right] {} (2)
    (4) edge[-] [left] node [left] {} (3);
\end{tikzpicture}
};
\draw[->, >=stealth', very thick]{} (a) -- (b);
\end{tikzpicture}

\end{center}
\caption{A diagrammatic representation of the $\mathbb{Z}_2^{\Omega'}\times \mathbb{Z}_4$ orientifold. Performing the $\mathbb{Z}_2^{\Omega'}$ orientifold with the representation $\gamma_{\Omega'}$ as in equation (\ref{FoldingEdgeEdge}) corresponds to folding the quiver across edges. Opposite nodes and edges are glued together, and the edges that are folded become anti-symmetric tensors, resulting in the quiver on the right side of the figure.}
\label{FoldingEdgeEdgeFig}
\end{figure}

For $\Gamma = \mathbb{Z}_2^{\Omega'} \times \mathbb{Z}_k$, when $k$ is even we have two choices for representations of $\Omega'$ on the Chan-Paton factors.
The first choice is:

\beq\label{FoldingEdgeEdge}
\gamma_{\Omega'} = \left(\begin{array}{cccccc}
    0 & 0 & 0 & 0 & 0 & -\mathds{1}_{m_0}\\
    0 & 0 & 0 & 0 & \iddots & 0 \\
    0 & 0 & 0 & -\mathds{1}_{m_{k/2-1}} & 0 & 0 \\
    0 & 0 & \mathds{1}_{m_{k/2}} & 0 & 0 & 0 \\
    0 & \iddots & 0 & 0 & 0 & 0 \\
    \mathds{1}_{m_{k-1}} & 0 & 0 & 0 & 0 & 0
\end{array} \right)
\eeq
where for consistency of this construction, we require that $m_i = m_{k-1-i}$.
With this choice, the $\mathbb{Z}_2^{\Omega'}$ orientifold identifies the $\U(m_i)$ vector multiplet with the $\U(m_{k-1-i})$ vector multiplet by imposing:

\beq
A_{\mu}^{(m_{k-1-i})} = - \left(A_{\mu}^{(m_{i})} \right)^{\intercal}
\eeq
The orientifold also identifies the corresponding bifundamental hypermultiplets, resulting in bifundamentals connecting all of the $U(m_i)$ vector multiplets and a hypermultiplet in the anti-symmetric tensor representation of $\U(m_0)$ and another in the same representation of $\U(m_{k/2-1})$.
This construction can be summarized as ``folding'' the $\mathbb{Z}_k$ quiver diagram, as shown in Figure \ref{FoldingEdgeEdgeFig}.

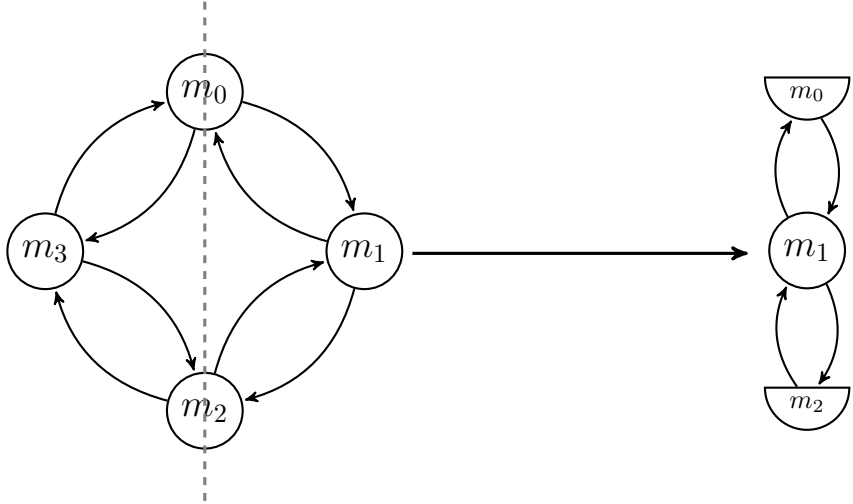
\begin{figure}
\begin{center}
\begin{tikzpicture}
\node[] (a) at (0,0) {
\begin{tikzpicture}[->,>=stealth',shorten >=1pt,auto,node distance=3cm,
                    thick,main node/.style={circle,draw,font=\sffamily\Large\bfseries}]

  \node[main node] (1) {$m_0$};
  \node[main node] (2) [below right of=1] {$m_1$};
  \node[main node] (3) [below left of=2] {$m_2$};
  \node[main node] (4) [below left of=1] {$m_3$};

  \draw[dashed, -, very thick, black!50] (0,1.2) -- (0,-5.5);
  \path[every node/.style={font=\sffamily\small}]
    (1) edge [bend left] node [left] {} (4)
        edge [bend left] node[left] {} (2)
    (2) edge [bend left] node [right] {} (1)
        edge [bend left] node [right] {} (3)
    (3) edge [bend left] node [right] {} (2)
        edge [bend left] node [right] {} (4)
    (4) edge [bend left] node [left] {} (3)
        edge [bend left] node[left] {} (1);
\end{tikzpicture}
};
\node[] (b) at (8,0) {
\begin{tikzpicture}[->,>=stealth',shorten >=1pt,auto,node distance=3cm,
                    thick,main node/.style={circle,draw,font=\sffamily\Large\bfseries},square node/.style={rectangle,draw},sp node/.style={shape=semicircle,draw, shape border rotate=180}]

  \node[sp node] (1) {$m_0$};
  \node[main node] (2) [below of=1, above=0.4cm] {$m_1$};
  \node[sp node] (3) [below of=2, above=0.6cm] {$m_2$};

  \path[every node/.style={font=\sffamily\small}]
    (1) edge [bend left] node [left] {} (2)
    (2) edge [bend left] node [right] {} (3)
    (2) edge [bend left] node [left] {} (1)
    (3) edge [bend left] node [right] {} (2);
\end{tikzpicture}
};
\draw[->, >=stealth', very thick]{} (a) -- (b);
\end{tikzpicture}

\end{center}
\caption{A diagrammatic representation of another $\mathbb{Z}_2^{\Omega'}\times \mathbb{Z}_4$ orientifold. Performing the $\mathbb{Z}_2^{\Omega'}$ orientifold with the representation $\tilde{\gamma}_{\Omega'}$ as in equation (\ref{FoldingNodeNode}) corresponds to folding the quiver across nodes. Opposite nodes and edges are glued together, and the nodes that are folded into themselves become $\USp$ gauge group factors, resulting in the quiver on the right side of the figure.}
\label{FoldingNodeNodeFig}
\end{figure}

The second possibility is:

\beq
\tilde{\gamma}_{\Omega'} = \left(\begin{array}{cccccccc}
    \Omega_{m_0} & 0 & 0 & 0 & 0 & 0 & 0 & 0\\
    0 & 0 & 0 & 0 & 0 & 0 & 0 & -\mathds{1}_{m_1}\\
    0 & 0 & 0 & 0 & 0 & 0 & \iddots & 0 \\
    0 & 0 & 0 & 0 & 0 & -\mathds{1}_{m_{k/2-1}} & 0 & 0 \\
    0 & 0 & 0 & 0 & \Omega_{m_{k/2}} & 0 & 0 & 0 \\
    0 & 0 & 0 & \mathds{1}_{m_{k/2+1}} & 0 & 0 & 0 & 0 \\
    0 & 0 & \iddots & 0 & 0 & 0 & 0 & 0 \\
    0 & \mathds{1}_{m_{k-1}} & 0 & 0 & 0 & 0 & 0 & 0
\end{array} \right)
\label{FoldingNodeNode}
\eeq
where we require that $m_i = m_{k-i}$ for $1 \le i \le k-1$, and

\beq
\Omega_{m} = \left(\begin{array}{cc}
     0 & -\mathds{1}_{m/2} \\
     \mathds{1}_{m/2} & 0
\end{array}\right).
\eeq

This orientifold acts similarly to the previous one on the multiplets transforming under gauge groups $\U(m_1)$ through $\U(m_{k/2-1})$, identifying them with those transforming under $\U(m_{k-1})$ through $\U(m_{k/2+1})$.
However, it acts differently on $\U(m_0)$ and $\U(m_{k/2})$.
Rather than identifying these with each other, it instead imposes conditions on the corresponding vector multiplets:
\beq
\left(\Omega_{m_0} A_{\mu}^{(m_0)}\right)^{\intercal} = -\Omega_{m_0} A_{\mu}^{(m_0)}
\eeq
So it changes these gauge group factors $\U(m_0), \U(m_{k/2}) \rightarrow \USp(m_0), \USp(m_{k/2})$.
We can represent this construction as another way of ``folding'' the $\mathbb{Z}_k$ quiver, see Figure \ref{FoldingNodeNodeFig}. We use \begin{tikzpicture}[so node/.style={shape=semicircle,draw}]
    \node[so node] (1) {m};
\end{tikzpicture} and \begin{tikzpicture}[sp node/.style={shape=semicircle,draw, shape border rotate=180}]
    \node[sp node] (1) {m};
\end{tikzpicture} to represent $\SO(m)$ and $\USp(m)$ gauge group factors respectively.

\subsubsection*{$k$ Odd}

For $\Gamma = \mathbb{Z}_2^{\Omega'} \times \mathbb{Z}_k$, when $k$ is odd, we have only one option:

\beq
\gamma_{\Omega'} = \left(\begin{array}{cccccccc}
    \Omega_{m_0} & 0 & 0 & 0 & 0 & 0 & 0\\
    0 & 0 & 0 & 0 & 0 & 0 & -\mathds{1}_{m_1}\\
    0 & 0 & 0 & 0 & 0 & \iddots & 0 \\
    0 & 0 & 0 & 0 & -\mathds{1}_{m_{(k-1)/2}} & 0 & 0 \\
    0 & 0 & 0 & \mathds{1}_{m_{(k-1)/2+1}} & 0 & 0 & 0 \\
    0 & 0 & \iddots & 0 & 0 & 0 & 0 \\
    0 & \mathds{1}_{m_{k-1}} & 0 & 0 & 0 & 0 & 0
\end{array} \right)
\label{FoldingEdgeNode}
\eeq

\begin{figure}
\begin{center}
\begin{tikzpicture}
\node[] (a) at (0,0) {
\begin{tikzpicture}[->,>=stealth',shorten >=1pt,auto,node distance=5cm,
                    thick,main node/.style={circle,draw,font=\sffamily\Large\bfseries}]

  \node[main node] (1) {$m_0$};
  \node[main node] (2) [below right of=1, left = 0.8cm] {$m_1$};
  \node[main node] (3) [below left of=1, right = 0.8cm] {$m_2$};

  \draw[dashed, -, very thick, black!50] (0,1.2) -- (0,-4.7);
  \path[every node/.style={font=\sffamily\small}]
    (1) edge [bend left=20] node [left] {} (3)
        edge [bend left=20] node[left] {} (2)
    (2) edge [bend left=20] node [right] {} (1)
        edge [bend left=20] node [right] {} (3)
    (3) edge [bend left=20] node [right] {} (2)
        edge [bend left=20] node [right] {} (1);
\end{tikzpicture}
};
\node[] (b) at (8,0) {
\begin{tikzpicture}[->,>=stealth',shorten >=1pt,auto,node distance=3cm,
                    thick,main node/.style={circle,draw,font=\sffamily\Large\bfseries},square node/.style={rectangle,draw},sp node/.style={shape=semicircle,draw, shape border rotate=180}]

  \node[sp node] (1) {$m_0$};
  \node[main node] (2) [below of=1] {$m_1$};
  \node[square node] (3) [below of=2,above=1cm] {$\begin{array}{c} \text{1 Anti-} \\
  \text{symmetric}
  \end{array}$};

  \path[every node/.style={font=\sffamily\small}]
    (1) edge [bend left] node [left] {} (2)
    (2) edge [bend left] node [left] {} (1)
    (2) edge[-] [left] node [right] {} (3);
\end{tikzpicture}
};
\draw[->, >=stealth', very thick]{} (a) -- (b);
\end{tikzpicture}

\end{center}
\caption{A diagrammatic representation of the $\mathbb{Z}_2^{\Omega'}\times \mathbb{Z}_3$ orientifold. Performing the $\mathbb{Z}_2^{\Omega'}$ orientifold with the representation $\gamma_{\Omega'}$ as in equation (\ref{FoldingEdgeNode}) corresponds to folding the quiver. Opposite nodes and edges are glued together. The node that is folded into itself becomes a $\USp$ gauge group factor, and the edges that are folded become an anti-symmetric tensor representation, resulting in the quiver on the right side of the figure.}
\label{FoldingEdgeNodeFig}
\end{figure}
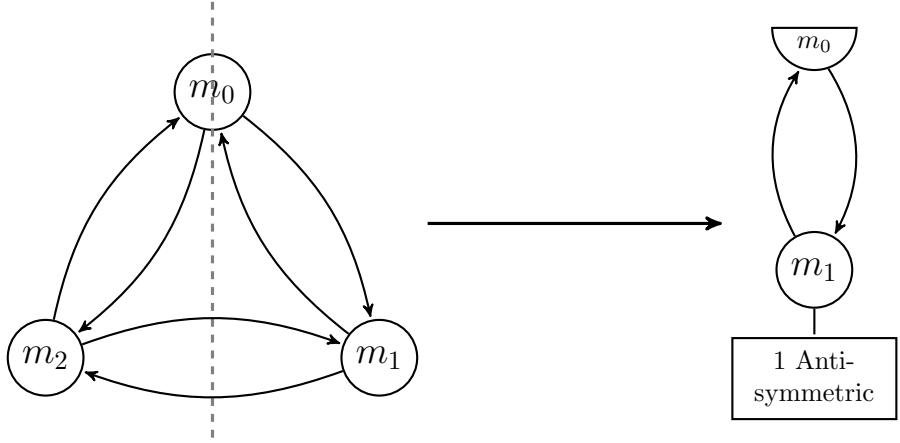

With this choice, performing the $\mathbb{Z}_2^{\Omega'}$ orientifold identifies the gauge group factors from $\U(m_1)$ to $\U(m_{(k-1)/2})$ with those from $\U(m_{k-1})$ to $\U(m_{(k-1)/2+1})$. It also changes $\U(m_0) \rightarrow \USp(m_0)$ and creates an antisymmetric tensor representation of $\U(m_{(k-1)/2})$.
This construction is summarized diagrammatically in Figure \ref{FoldingEdgeNodeFig}.

\subsection{$\mathbb{Z}_2^{\Omega'} \times \Dic_k$}

Next, let's consider the orientifold $\mathbb{Z}_2^{\Omega'}\times \Dic_k$, where the orientation preserving subgroup $\Dic_k$ is the binary dicyclic group of order $4k$.
The binary dicyclic group $\Dic_k = \left<\alpha, j|\alpha^{2k} = 1, j^2 = \alpha^k, \alpha j = j \alpha^{-1} \right>$ has two generators, and we must choose how they act on the Chan-Paton factors.
Up to a change of basis, we can always represent the generators as:

\beq
\gamma_{\alpha} = \left(\begin{array}{cccccccc}
    \mathds{1}_{m_0+m_1} & 0 & 0 & 0 & 0 & 0 & 0 & 0 \\
    0 & \omega_{2k} \mathds{1}_{m_2} & 0 & 0 & 0 & 0 & 0 & 0 \\
    0 & 0 & \ddots & 0 & 0 & 0 & 0 & 0 \\
    0 & 0 & 0 & \omega_{2k}^{k-1} \mathds{1}_{m_k} & 0 & 0 & 0 & 0 \\
    0 & 0 & 0 & 0 & \omega_{2k}^{k} \mathds{1}_{m_{k+1} + m_{k+2}} & 0 & 0 & 0\\
    0 & 0 & 0 & 0 & 0 & \omega_{2k}^{k+1} \mathds{1}_{m_k} & 0 & 0 \\
    0 & 0 & 0 & 0 & 0 & 0 & \ddots & 0 \\
    0 & 0 & 0 & 0 & 0 & 0 & 0 & \omega_{2k}^{2k-1} \mathds{1}_{m_2}
\end{array} \right)
\eeq
and
\beq
\gamma_j = \left(\begin{array}{ccccccccccc}
    \mathds{1}_{m_0} & 0 & 0 & 0 & 0 & 0 & 0 & 0 & 0 & 0 & 0 \\
    0 & -\mathds{1}_{m_1} & 0 & 0 & 0 & 0 & 0 & 0 & 0 & 0 & 0 \\
    0 & 0 & 0 & 0 & 0 & 0 & 0 & 0 & 0 & 0 & -\mathds{1}_{m_2} \\
    0 & 0 & 0 & 0 & 0 & 0 & 0 & 0 & 0 & \mathds{1}_{m_3} & 0 \\
    0 & 0 & 0 & 0 & 0 & 0 & 0 & 0 & \iddots & 0 & 0 \\
    0 & 0 & 0 & 0 & 0 & 0 & 0 & (-1)^{k+1}\mathds{1}_{m_k} & 0 & 0 & 0 \\
    0 & 0 & 0 & 0 & 0 & -\mathds{1}_{m_{k+1}} & 0 & 0 & 0 & 0 & 0 \\
    0 & 0 & 0 & 0 & 0 & 0 & \mathds{1}_{m_{k+2}} & 0 & 0 & 0 & 0 \\
    0 & 0 & 0 & 0 & \mathds{1}_{m_k} & 0 & 0 & 0 & 0 & 0 & 0 \\
    0 & 0 & 0 & \iddots & 0 & 0 & 0 & 0 & 0 & 0 & 0 \\
    0 & 0 & \mathds{1}_{m_2} & 0 & 0 & 0 & 0 & 0 & 0 & 0 & 0
\end{array} \right).
\eeq

\begin{figure}
\begin{center}
\begin{tikzpicture}
\node[] (a) at (0,0) {
\begin{tikzpicture}[->,>=stealth',shorten >=1pt,auto,node distance=1.5cm,
                    thick,main node/.style={circle,draw,font=\sffamily\Large\bfseries}]

  \node[main node] (1) {$m_0$};
  \node[main node] (2) [below right of=1] {$m_2$};
  \node[main node] (3) [below left of=2] {$m_1$};
  \node[main node] (7) [right of=2] {$m_3$};
  \node[main node] (4) [right of=7]{$m_4$};
  \node[main node] (5) [above right of=4]{$m_6$};
  \node[main node] (6) [below right of=4]{$m_5$};

  \path[every node/.style={font=\sffamily\small}]
    (1) edge [bend left=20] node [left] {} (2)
    (2) edge [bend left=20] node [right] {} (1)
        edge [bend left=20] node [right] {} (3)
        edge [bend left=20] node [right] {} (7)
    (3) edge [bend left=20] node [right] {} (2)
    (7) edge [bend left=20] node [right] {} (4)
        edge [bend left=20] node [right] {} (2)
    (4) edge [bend left=20] node [right] {} (7)
        edge [bend left=20] node [right] {} (5)
        edge [bend left=20] node [right] {} (6)
    (5) edge [bend left=20] node [right] {} (4)
    (6) edge [bend left=20] node [right] {} (4);
\end{tikzpicture}
};
\node[] (b) at (8,0) {
\begin{tikzpicture}[->,>=stealth',shorten >=1pt,auto,node distance=1.5cm,
                    thick,main node/.style={circle,draw,font=\sffamily\Large\bfseries},sp node/.style={shape=semicircle,draw, shape border rotate=180},so node/.style={shape=semicircle,draw}]

  \node[sp node] (1) {$m_0$};
  \node[so node] (2) [below right of=1] {$m_2$};
  \node[sp node] (3) [below left of=2] {$m_1$};
  \node[sp node] (7) [right of=2] {$m_3$};
  \node[so node] (4) [right of=7]{$m_4$};
  \node[sp node] (5) [above right of=4]{$m_6$};
  \node[sp node] (6) [below right of=4]{$m_5$};

  \path[every node/.style={font=\sffamily\small}]
    (1) edge[-] node [left] {} (2)
    (2) edge[-] node [right] {} (7)
    (3) edge[-] node [left] {} (2)
    (7) edge[-] node [right] {} (4)
    (4) edge[-] node [right] {} (5)
        edge[-] node [right] {} (6);
\end{tikzpicture}
};
\node[] (aLabel) at (0,-1.5) {(a)};
\node[] (bLabel) at (8,-1.5) {(b)};
\draw[->, >=stealth', very thick]{} (a) -- (b);
\end{tikzpicture}
\end{center}
\caption{A diagrammatic representation of the $\mathbb{Z}_2^{\Omega'} \times \Dic_4$ orientifold with $\Omega'$ represented as in equation (\ref{AlternatingDQuiver}).}
\label{AlternatingDQuiverFig}
\end{figure}
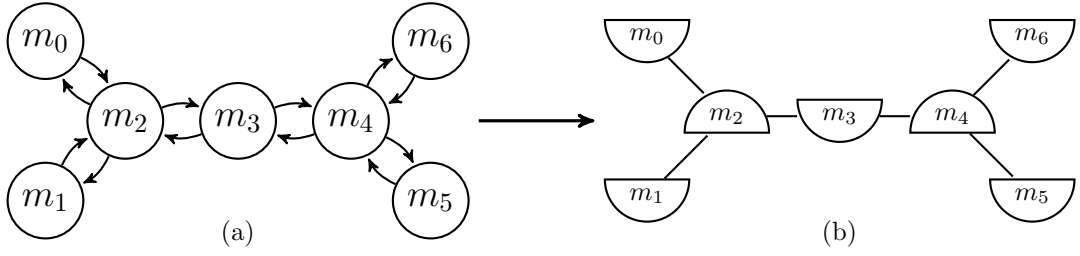

The orbifolded theory contains $k+2$, $\U(m_i)$ vectormultiplets, corresponding to the irreducible representations of $\Dic_k$.
In addition, it contains hypermultiplets transforming in $\textbf{fund}\times \overline{\textbf{fund}}$ representations of the $U(m_i)$ factors which are connected by edges in the affine $D_{k+2}$ Dynkin diagram, see Figure \ref{AlternatingDQuiverFig}(a).

To find the orientifolded theory, we must also impose invariance under the action of $\Omega'$.
First, we must choose how $\Omega'$ acts on the Chan-Paton factors.

\subsubsection*{$k$ Even}

For $\Gamma = \mathbb{Z}_2^{\Omega'} \times \Dic_k$, when $k$ is even, we may choose to represent $\Omega'$ as
\beq
\label{AlternatingDQuiver}
\gamma_{\Omega'} = \left(\begin{array}{cccccccccccc}
    \Omega_{m_0} & 0 & 0 & 0 & 0 & 0 & 0 & 0 & 0 & 0 & 0 & 0 \\
    0 & -\Omega_{m_1} & 0 & 0 & 0 & 0 & 0 & 0 & 0 & 0 & 0 & 0 \\
    0 & 0 & 0 & 0 & 0 & 0 & 0 & 0 & 0 & 0 & 0 & \mathds{1}_{m_2} \\
    0 & 0 & 0 & 0 & 0 & 0 & 0 & 0 & 0 & 0 & \Omega_{m_3} & 0 \\
    0 & 0 & 0 & 0 & 0 & 0 & 0 & 0 & 0 & \iddots & 0 & 0 \\
    0 & 0 & 0 & 0 & 0 & 0 & 0 & 0 & \mathds{1}_{m_k} & 0 & 0 & 0 \\
    0 & 0 & 0 & 0 & 0 & 0 & -\Omega_{m_{k+1}} & 0 & 0 & 0 & 0 & 0 \\
    0 & 0 & 0 & 0 & 0 & 0 & 0 & \Omega_{m_{k+2}} & 0 & 0 & 0 & 0 \\
    0 & 0 & 0 & 0 & 0 & -\mathds{1}_{m_k} & 0 & 0 & 0 & 0 & 0 & 0 \\
    0 & 0 & 0 & 0 & \iddots & 0 & 0 & 0 & 0 & 0 & 0 & 0 \\
    0 & 0 & 0 & \Omega_{m_3} & 0 & 0 & 0 & 0 & 0 & 0 & 0 & 0 \\
    0 & 0 & -\mathds{1}_{m_2} & 0 & 0 & 0 & 0 & 0 & 0 & 0 & 0 & 0
\end{array} \right).
\eeq
With this choice, the orientifold results in the affine $D$-type quiver with alternating $\SO$ and $\USp$ gauge group factors, see Figure \ref{AlternatingDQuiverFig}(b).

\begin{figure}
\begin{center}
\begin{tikzpicture}
\node[] (a) at (0,0) {
\begin{tikzpicture}[->,>=stealth',shorten >=1pt,auto,node distance=1.5cm,
                    thick,main node/.style={circle,draw,font=\sffamily\Large\bfseries}]

  \node[main node] (1) {$m_0$};
  \node[main node] (2) [below right of=1] {$m_2$};
  \node[main node] (3) [below left of=2] {$m_1$};
  \node[main node] (7) [right of=2] {$m_3$};
  \node[main node] (4) [right of=7]{$m_4$};
  \node[main node] (5) [above right of=4]{$m_6$};
  \node[main node] (6) [below right of=4]{$m_5$};

  \path[every node/.style={font=\sffamily\small}]
    (1) edge [bend left=20] node [left] {} (2)
    (2) edge [bend left=20] node [right] {} (1)
        edge [bend left=20] node [right] {} (3)
        edge [bend left=20] node [right] {} (7)
    (3) edge [bend left=20] node [right] {} (2)
    (7) edge [bend left=20] node [right] {} (4)
        edge [bend left=20] node [right] {} (2)
    (4) edge [bend left=20] node [right] {} (7)
        edge [bend left=20] node [right] {} (5)
        edge [bend left=20] node [right] {} (6)
    (5) edge [bend left=20] node [right] {} (4)
    (6) edge [bend left=20] node [right] {} (4);
\end{tikzpicture}
};
\node[] (b) at (8,0) {
\begin{tikzpicture}[->,>=stealth',shorten >=1pt,auto,node distance=1.5cm,
                    thick,main node/.style={circle,draw,font=\sffamily\Large\bfseries},sp node/.style={shape=semicircle,draw, shape border rotate=180},so node/.style={shape=semicircle,draw}]

  \node[main node] (1) {$m_0$};
  \node[sp node] (2) [right of=1] {$m_2$};
  \node[so node] (7) [right of=2] {$m_3$};
  \node[sp node] (4) [right of=7]{$m_4$};
  \node[main node] (5) [right of=4]{$m_6$};

  \path[every node/.style={font=\sffamily\small}]
    (1) edge [bend left] node [left] {} (2)
    (2) edge[-] node [right] {} (7)
        edge [bend left] node [left] {} (1)
    (7) edge[-] node [right] {} (4)
    (4) edge [bend left] node [right] {} (5)
    (5) edge [bend left] node [right] {} (4);
\end{tikzpicture}
};
\draw[->, >=stealth', very thick]{} (a) -- (b);
\end{tikzpicture}
\end{center}
\caption{A diagrammatic representation of the $\mathbb{Z}_2^{\Omega'} \times \Dic_4$ orientifold with $\Omega'$ represented as in equation (\ref{SOUSpLinear}).}
\label{SOUSpLinearFig}
\end{figure}
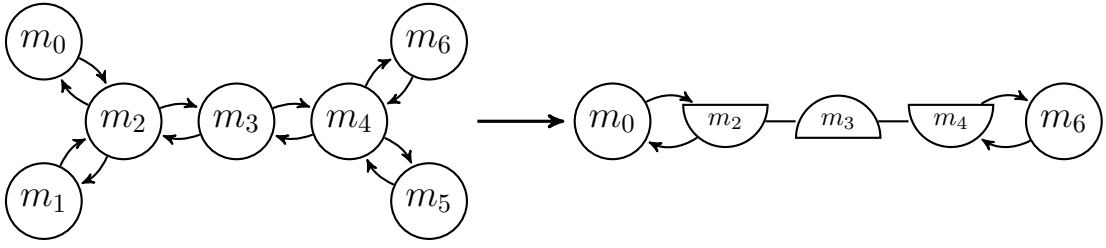

Another choice for $\Gamma = \mathbb{Z}_2^{\Omega'} \times \Dic_k$, when $k$ is even, when $m_0=m_1$ and $m_{k+1} = m_{k+2}$ is
\beq
\label{SOUSpLinear}
\tilde{\gamma}_{\Omega'} = \left(\begin{array}{cccccccccccc}
    0 & -\mathds{1}_{m_0} & 0 & 0 & 0 & 0 & 0 & 0 & 0 & 0 & 0 & 0 \\
    \mathds{1}_{m_1} & 0 & 0 & 0 & 0 & 0 & 0 & 0 & 0 & 0 & 0 & 0 \\
    0 & 0 & 0 & 0 & 0 & 0 & 0 & 0 & 0 & 0 & 0 & \Omega_{m_2} \\
    0 & 0 & 0 & 0 & 0 & 0 & 0 & 0 & 0 & 0 & \mathds{1}_{m_3} & 0 \\
    0 & 0 & 0 & 0 & 0 & 0 & 0 & 0 & 0 & \iddots & 0 & 0 \\
    0 & 0 & 0 & 0 & 0 & 0 & 0 & 0 & \Omega_{m_k} & 0 & 0 & 0 \\
    0 & 0 & 0 & 0 & 0 & 0 & 0 & \mathds{1}_{m_{k+1}} & 0 & 0 & 0 & 0 \\
    0 & 0 & 0 & 0 & 0 & 0 & -\mathds{1}_{m_{k+2}} & 0 & 0 & 0 & 0 & 0 \\
    0 & 0 & 0 & 0 & 0 & \Omega_{m_k} & 0 & 0 & 0 & 0 & 0 & 0 \\
    0 & 0 & 0 & 0 & \iddots & 0 & 0 & 0 & 0 & 0 & 0 & 0 \\
    0 & 0 & 0 & -\mathds{1}_{m_3} & 0 & 0 & 0 & 0 & 0 & 0 & 0 & 0 \\
    0 & 0 & \Omega_{m_2} & 0 & 0 & 0 & 0 & 0 & 0 & 0 & 0 & 0
\end{array} \right).
\eeq
With this choice we get a linear quiver of alternating $\SO$ and $\USp$ gauge group factors, with a $\U(m_0)$ factor on one end of the quiver, and a $\U(m_{k+2})$ factor on the other end, see Figure \ref{SOUSpLinearFig}.

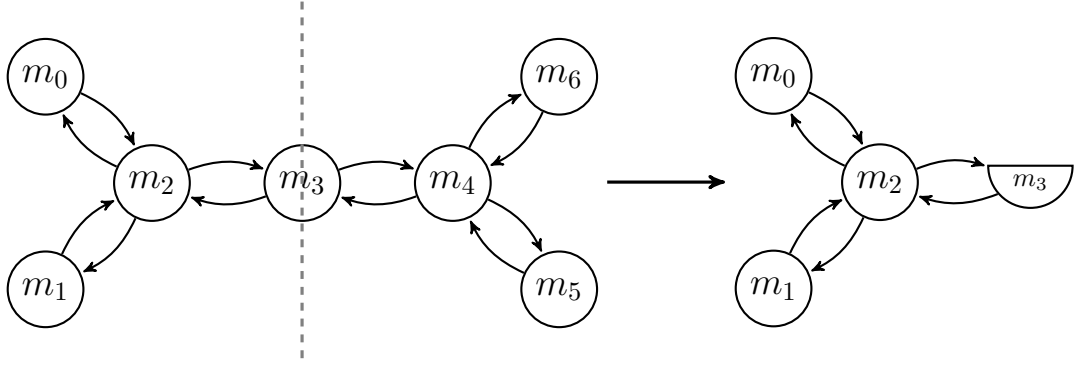
\begin{figure}
\begin{center}
\begin{tikzpicture}
\node[] (a) at (0,0) {
\begin{tikzpicture}[->,>=stealth',shorten >=1pt,auto,node distance=2cm,
                    thick,main node/.style={circle,draw,font=\sffamily\Large\bfseries}]

  \node[main node] (1) {$m_0$};
  \node[main node] (2) [below right of=1] {$m_2$};
  \node[main node] (3) [below left of=2] {$m_1$};
  \node[main node] (7) [right of=2] {$m_3$};
  \node[main node] (4) [right of=7]{$m_4$};
  \node[main node] (5) [above right of=4]{$m_6$};
  \node[main node] (6) [below right of=4]{$m_5$};

  \draw[dashed, -, very thick, black!50] (3.41,1) -- (3.41,-3.8);
  \path[every node/.style={font=\sffamily\small}]
    (1) edge [bend left=20] node [left] {} (2)
    (2) edge [bend left=20] node [right] {} (1)
        edge [bend left=20] node [right] {} (3)
        edge [bend left=20] node [right] {} (7)
    (3) edge [bend left=20] node [right] {} (2)
    (7) edge [bend left=20] node [right] {} (4)
        edge [bend left=20] node [right] {} (2)
    (4) edge [bend left=20] node [right] {} (7)
        edge [bend left=20] node [right] {} (5)
        edge [bend left=20] node [right] {} (6)
    (5) edge [bend left=20] node [right] {} (4)
    (6) edge [bend left=20] node [right] {} (4);
\end{tikzpicture}
};
\node[] (b) at (8,0) {
\begin{tikzpicture}[->,>=stealth',shorten >=1pt,auto,node distance=2cm,
                    thick,main node/.style={circle,draw,font=\sffamily\Large\bfseries},square node/.style={rectangle,draw},sp node/.style={shape=semicircle,draw, shape border rotate=180}]

  \node[main node] (1) {$m_0$};
  \node[main node] (2) [below right of=1] {$m_2$};
  \node[main node] (3) [below left of=2] {$m_1$};
  \node[sp node] (4) [right of=2] {$m_3$};

  \path[every node/.style={font=\sffamily\small}]
    (1) edge [bend left=20] node [left] {} (2)
    (2) edge [bend left=20] node [left] {} (1)
        edge [bend left=20] node [left] {} (3)
        edge [bend left=20] node [left] {} (4)
    (3) edge [bend left=20] node [right] {} (2)
    (4) edge [bend left=20] node [right] {} (2);
\end{tikzpicture}
};
\draw[->, >=stealth', very thick]{} (a) -- (b);
\end{tikzpicture}
\end{center}
\caption{A diagrammatic representation of a $\mathbb{Z}_2^{\Omega'} \times \Dic_4$ orientifold.}
\label{FoldingDicNodeFig}
\end{figure}

Another choice for $\Gamma = \mathbb{Z}_2^{\Omega'} \times \Dic_k$, of a somewhat different nature, when $k$ is even, is given by:
\beq
\gamma_{\Omega'} = \left(\begin{array}{cc}
    \sigma_{\Omega'} & 0 \\
    0 & \sigma_{\Omega'}
    \end{array}
\right)
\eeq
where
\beq\label{FoldingDicNode}
\sigma_{\Omega'} = \left(\begin{array}{ccccccccccc}
    0 & 0 & 0 & 0 & 0 & 0 & -\mathds{1}_{m_0} \\
    0 & 0 & 0 & 0 & 0 & \iddots & 0 \\
    0 & 0 & 0 & 0 & -\mathds{1}_{m_{k/2}} & 0 & 0 \\
    0 & 0 & 0 & \Omega_{m_{k/2+1}} & 0 & 0 & 0 \\
    0 & 0 & \mathds{1}_{m_{k/2 + 2}} & 0 & 0 & 0 & 0 \\
    0 & \iddots & 0 & 0 & 0 & 0 & 0 \\
    \mathds{1}_{m_{k+2}} & 0 & 0 & 0 & 0 & 0 & 0
\end{array} \right).
\eeq
where we assume we have $m_{k+2-i} = m_i$.
With this choice, the orientifold identifies $U(m_i)$ gauge group factors in pairs, except at the center of the quiver, where a node is ``folded'' into a $\USp(m_{k/2+1})$ factor.
This is summarized in Figure \ref{FoldingDicNodeFig}.

\subsubsection*{$k$ Odd}

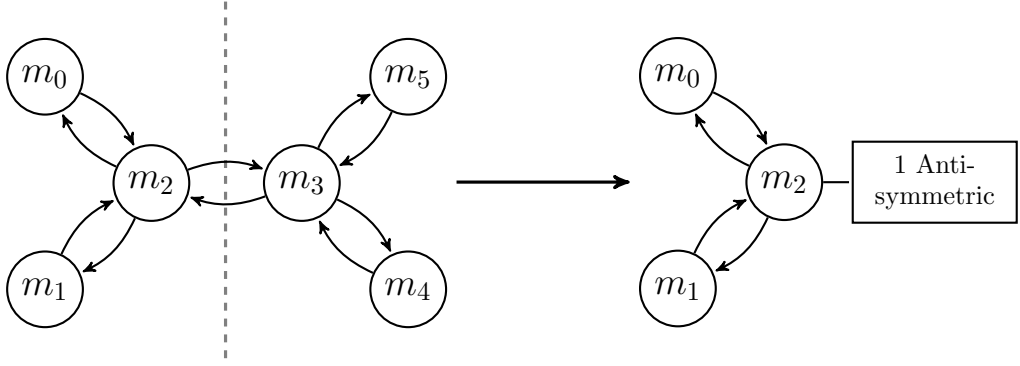
\begin{figure}
\begin{center}
\begin{tikzpicture}
\node[] (a) at (0,0) {
\begin{tikzpicture}[->,>=stealth',shorten >=1pt,auto,node distance=2cm,
                    thick,main node/.style={circle,draw,font=\sffamily\Large\bfseries}]

  \node[main node] (1) {$m_0$};
  \node[main node] (2) [below right of=1] {$m_2$};
  \node[main node] (3) [below left of=2] {$m_1$};
  \node[main node] (4) [right of=2]{$m_3$};
  \node[main node] (5) [above right of=4]{$m_5$};
  \node[main node] (6) [below right of=4]{$m_4$};

  \draw[dashed, -, very thick, black!50] (2.4,1) -- (2.4,-3.8);
  \path[every node/.style={font=\sffamily\small}]
    (1) edge [bend left=20] node [left] {} (2)
    (2) edge [bend left=20] node [right] {} (1)
        edge [bend left=20] node [right] {} (3)
        edge [bend left=20] node [right] {} (4)
    (3) edge [bend left=20] node [right] {} (2)
    (4) edge [bend left=20] node [right] {} (2)
        edge [bend left=20] node [right] {} (5)
        edge [bend left=20] node [right] {} (6)
    (5) edge [bend left=20] node [right] {} (4)
    (6) edge [bend left=20] node [right] {} (4);
\end{tikzpicture}
};
\node[] (b) at (8,0) {
\begin{tikzpicture}[->,>=stealth',shorten >=1pt,auto,node distance=2cm,
                    thick,main node/.style={circle,draw,font=\sffamily\Large\bfseries},square node/.style={rectangle,draw},sp node/.style={shape=semicircle,draw, shape border rotate=180}]

  \node[main node] (1) {$m_0$};
  \node[main node] (2) [below right of=1] {$m_2$};
  \node[main node] (3) [below left of=2] {$m_1$};
  \node[square node] (4) [right of=2] {$\begin{array}{c} \text{1 Anti-} \\
  \text{symmetric}
  \end{array}$};

  \path[every node/.style={font=\sffamily\small}]
    (1) edge [bend left=20] node [left] {} (2)
    (2) edge [bend left=20] node [left] {} (1)
        edge [bend left=20] node [left] {} (3)
        edge[-] node [left] {} (4)
    (3) edge [bend left=20] node [right] {} (2);
\end{tikzpicture}
};
\draw[->, >=stealth', very thick]{} (a) -- (b);
\end{tikzpicture}
\end{center}
\caption{A diagrammatic representation of a $\mathbb{Z}_2^{\Omega'} \times \Dic_3$ orientifold.}
\label{FoldingDicEdgeFig}
\end{figure}

For $\Gamma = \mathbb{Z}_2^{\Omega'} \times \Dic_k$, when $k$ is odd, assuming we have $m_{k+2-i} = m_i$, we can choose:
\beq
\gamma_{\Omega'} = \left(\begin{array}{cc}
    \sigma_{\Omega'} & 0 \\
    0 & \sigma_{\Omega'}
    \end{array}
\right)
\eeq
where
\beq\label{FoldingDicEdge}
\sigma_{\Omega'} = \left(\begin{array}{cccccccccc}
    0 & 0 & 0 & 0 & 0 & -\mathds{1}_{m_0} \\
    0 & 0 & 0 & 0 & \iddots & 0 \\
    0 & 0 & 0 & -\mathds{1}_{m_{(k+1)/2}} & 0 & 0 \\
    0 & 0 & \mathds{1}_{m_{(k+3)/2}} & 0 & 0 & 0 \\
    0 & \iddots & 0 & 0 & 0 & 0 \\
    \mathds{1}_{m_{k+2}} & 0 & 0 & 0 & 0 & 0
\end{array} \right).
\eeq
With this choice, the orientifold identifies $U(m_i)$ gauge group factors in pairs, and produces an anti-symmetric tensor representation of $U(m_{(k+1)/2})$, see Figure \ref{FoldingDicEdgeFig}.

\subsection{$\mathbb{Z}_{2k}^{\alpha \Omega'}$}

Now let's consider the orientifold by $\mathbb{Z}_{2k}^{\alpha \Omega'} = \left<\alpha \Omega'| \alpha \in \SU(2), \alpha^{2k} = 1\right>$.

\subsubsection*{$k$ Even}

For $\Gamma= \mathbb{Z}_{2k}^{\alpha \Omega'}$, when $k$ is even, one way of representing the generator on the Chan-Paton factors is
\beq
\label{ZkOrientEdgeEdge}
\gamma_{\alpha \Omega'} = \left(\begin{array}{ccccccc}
    0 & \omega_{2k} \mathds{1}_{m_0} & 0 & 0 & 0 & 0 & 0 \\
    \mathds{1}_{m_0} & 0 & 0 & 0 & 0 & 0 & 0 \\
    0 & 0 & 0 & \omega_{2k}^3 \mathds{1}_{m_1} & 0 & 0 & 0 \\
    0 & 0 & \mathds{1}_{m_1} & 0 & 0 & 0 & 0 \\
    0 & 0 & 0 & 0 & \ddots & 0 & 0 \\
    0 & 0 & 0 & 0 & 0 & 0 & \omega_{2k}^{k-1} \mathds{1}_{m_{k/2-1}} \\
    0 & 0 & 0 & 0 & 0 & \mathds{1}_{m_{k/2-1}} & 0
\end{array} \right)
\eeq
This orientifold gives the linear quiver with $\U(m_i)$ gauge group factors, with a symmetric tensor representation on one end of the quiver and an anti-symmetric tensor representation on the other end.

\begin{figure}
\begin{center}
\begin{tikzpicture}
\node[] (a) at (0,0) {
\begin{tikzpicture}[->,>=stealth',shorten >=1pt,auto,node distance=3cm,
                    thick,main node/.style={circle,draw,font=\sffamily\Large\bfseries}]

  \node[main node] (1) {$m_0$};
  \node[main node] (2) [right of=1] {$m_0$};
  \node[main node] (3) [below of=2] {$m_1$};
  \node[main node] (4) [below of=1] {$m_1$};

  \draw[dashed, -, very thick, black!50] (1.5,1.2) -- (1.5,-4.2);
  
  \path[every node/.style={font=\sffamily\small}]
    (1) edge [bend left] node [left] {} (4)
        edge [bend left] node[left] {} (2)
    (2) edge [bend left] node [right] {} (1)
        edge [bend left] node [right] {} (3)
    (3) edge [bend left] node [right] {} (2)
        edge [bend left] node [right] {} (4)
    (4) edge [bend left] node [left] {} (3)
        edge [bend left] node[left] {} (1);
        
\end{tikzpicture}
};
\node[] (b) at (8,0) {
\begin{tikzpicture}[->,>=stealth',shorten >=1pt,auto,node distance=3cm,
                    thick,main node/.style={circle,draw,font=\sffamily\Large\bfseries},square node/.style={rectangle,draw}]

  \node[square node] (1) {$\begin{array}{c} \text{1 Anti-} \\
  \text{symmetric}
  \end{array}$};
  \node[main node] (2) [below of=1,above=1cm] {$m_0$};
  \node[main node] (3) [below of=2] {$m_1$};
  \node[square node] (4) [below of=3,above=1.5cm] {1 Symmetric};

  \path[every node/.style={font=\sffamily\small}]
    (1) edge[-] [left] node [left] {} (2)
    (2) edge [bend left] node [right] {} (3)
    (3) edge [bend left] node [right] {} (2)
    (4) edge[-] [left] node [left] {} (3);
\end{tikzpicture}
};
\draw[->, >=stealth', very thick]{} (a) -- (b);
\end{tikzpicture}

\end{center}
\caption{A diagrammatic representation of a $\mathbb{Z}_8^{\alpha\Omega'}$ orientifold. The orientation-preserving subgroup is $G_1 = \mathbb{Z}_4 = \langle \alpha^2 \rangle$. Performing the further orientifold by $\alpha \Omega'$ with the representation $\gamma_{\alpha \Omega'}$ as in equation (\ref{ZkOrientEdgeEdge}) corresponds to folding the quiver across edges. Opposite nodes and edges are glued together, and the edges that are folded become a symmetric and an anti-symmetric tensor, resulting in the quiver on the right side of the figure.}
\label{ZkOrientEdgeEdgeFig}
\end{figure}
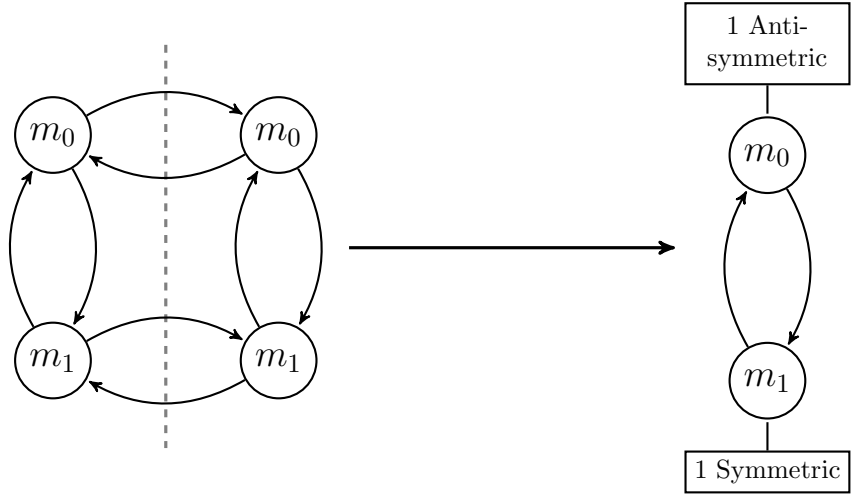

Another choice of representation is given by
\beq
\label{ZkOrientNodeNode}
\tilde{\gamma}_{\alpha \Omega'} = \left(\begin{array}{ccccccc}
    0 & \mathds{1}_{m_0} & 0 & 0 & 0 & 0 & 0 \\
    \mathds{1}_{m_0} & 0 & 0 & 0 & 0 & 0 & 0 \\
    0 & 0 & 0 & \omega_{2k}^2 \mathds{1}_{m_1} & 0 & 0 & 0 \\
    0 & 0 & \mathds{1}_{m_1} & 0 & 0 & 0 & 0 \\
    0 & 0 & 0 & 0 & \ddots & 0 & 0 \\
    0 & 0 & 0 & 0 & 0 & 0 & -\mathds{1}_{m_{k/2}} \\
    0 & 0 & 0 & 0 & 0 & \mathds{1}_{m_{k/2}} & 0
\end{array} \right)
\eeq
This orientifold gives the linear quiver with $\U(m_i)$ gauge group factors, with an $\SO(2 m_0)$ factor on one end of the quiver and a $\USp(2 m_{k/2})$ factor on the other end.

\begin{figure}
\begin{center}
\begin{tikzpicture}
\node[] (a) at (0,0) {
\begin{tikzpicture}[->,>=stealth',shorten >=1pt,auto,node distance=3cm,
                    thick,main node/.style={circle,draw,font=\sffamily\Large\bfseries}]

  \node[main node] (1) {$2m_0$};
  \node[main node] (2) [below right of=1] {$m_1$};
  \node[main node] (3) [below left of=2] {$2m_2$};
  \node[main node] (4) [below left of=1] {$m_1$};

  \draw[dashed, -, very thick, black!50] (0,1.2) -- (0,-5.5);
  \path[every node/.style={font=\sffamily\small}]
    (1) edge [bend left] node [left] {} (4)
        edge [bend left] node[left] {} (2)
    (2) edge [bend left] node [right] {} (1)
        edge [bend left] node [right] {} (3)
    (3) edge [bend left] node [right] {} (2)
        edge [bend left] node [right] {} (4)
    (4) edge [bend left] node [left] {} (3)
        edge [bend left] node[left] {} (1);
\end{tikzpicture}
};
\node[] (b) at (8,0) {
\begin{tikzpicture}[->,>=stealth',shorten >=1pt,auto,node distance=3cm,
                    thick,main node/.style={circle,draw,font=\sffamily\Large\bfseries},square node/.style={rectangle,draw},sp node/.style={shape=semicircle,draw, shape border rotate=180},so node/.style={shape=semicircle,draw}]

  \node[so node] (1) {$2m_0$};
  \node[main node] (2) [below of=1, above=0.4cm] {$m_1$};
  \node[sp node] (3) [below of=2, above=0.6cm] {$2m_2$};

  \path[every node/.style={font=\sffamily\small}]
    (1) edge [bend left] node [left] {} (2)
    (2) edge [bend left] node [right] {} (3)
    (2) edge [bend left] node [left] {} (1)
    (3) edge [bend left] node [right] {} (2);
\end{tikzpicture}
};
\draw[->, >=stealth', very thick]{} (a) -- (b);
\end{tikzpicture}

\end{center}
\caption{A diagrammatic representation of another $\mathbb{Z}_8^{\alpha \Omega'}$ orientifold. The orientation preserving subgroup is $G_1=\mathbb{Z}_4 = \langle\alpha^2\rangle$. Performing the further orientifold by $\alpha \Omega'$, with the representation $\tilde{\gamma}_{\alpha \Omega'}$ as in equation (\ref{ZkOrientNodeNode}) corresponds to folding the quiver across nodes. Opposite nodes and edges are glued together, and the nodes that are folded into themselves become an $\USp$ and an $\SO$ gauge group factor, resulting in the quiver on the right side of the figure.}
\label{ZkOrientNodeNodeFig}
\end{figure}
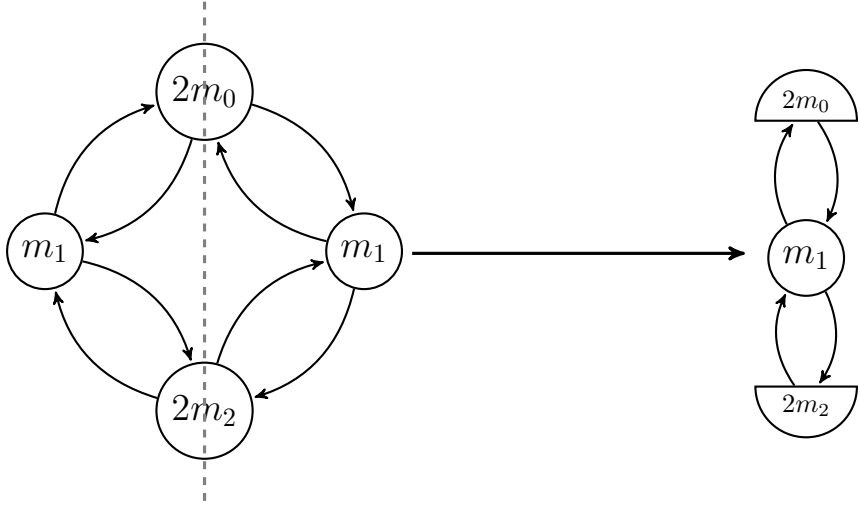

\subsubsection*{$k$ Odd}

For $\Gamma= \mathbb{Z}_{2k}^{\alpha \Omega'}$, when k is odd, we also have two possible choices for representation.
The first choice is
\beq
\gamma_{\alpha \Omega'} = \left(\begin{array}{ccccccc}
    0 & \mathds{1}_{m_0} & 0 & 0 & 0 & 0 & 0 \\
    \mathds{1}_{m_0} & 0 & 0 & 0 & 0 & 0 & 0 \\
    0 & 0 & 0 & \omega_{2k}^2 \mathds{1}_{m_1} & 0 & 0 & 0 \\
    0 & 0 & \mathds{1}_{m_1} & 0 & 0 & 0 & 0 \\
    0 & 0 & 0 & 0 & \ddots & 0 & 0 \\
    0 & 0 & 0 & 0 & 0 & 0 & \omega_{2k}^{k-1} \mathds{1}_{m_{(k-1)/2}} \\
    0 & 0 & 0 & 0 & 0 & \mathds{1}_{m_{(k-1)/2}} & 0
\end{array} \right)
\eeq
The orientifold with this choice of representation results in a linear quiver with $\U(m_i)$ gauge group factors, with an $\SO(2 m_0)$ factor on one end of the quiver and an anti-symmetric tensor representation on the other end.

\begin{figure}
\begin{center}
\begin{tikzpicture}
\node[] (a) at (0,0) {
\begin{tikzpicture}[->,>=stealth',shorten >=1pt,auto,node distance=5cm,
                    thick,main node/.style={circle,draw,font=\sffamily\Large\bfseries}]

  \node[main node] (1) {$2m_0$};
  \node[main node] (2) [below right of=1, left = 0.8cm] {$m_1$};
  \node[main node] (3) [below left of=1, right = 0.8cm] {$m_1$};

  \draw[dashed, -, very thick, black!50] (0,1.2) -- (0,-4.7);
  \path[every node/.style={font=\sffamily\small}]
    (1) edge [bend left=20] node [left] {} (3)
        edge [bend left=20] node[left] {} (2)
    (2) edge [bend left=20] node [right] {} (1)
        edge [bend left=20] node [right] {} (3)
    (3) edge [bend left=20] node [right] {} (2)
        edge [bend left=20] node [right] {} (1);
\end{tikzpicture}
};
\node[] (b) at (8,0) {
\begin{tikzpicture}[->,>=stealth',shorten >=1pt,auto,node distance=3cm,
                    thick,main node/.style={circle,draw,font=\sffamily\Large\bfseries},square node/.style={rectangle,draw},sp node/.style={shape=semicircle,draw, shape border rotate=180},so node/.style={shape=semicircle,draw}]

  \node[so node] (1) {$2m_0$};
  \node[main node] (2) [below of=1] {$m_1$};
  \node[square node] (3) [below of=2,above=1cm] {$\begin{array}{c}
       \text{1 Anti-}\\
       \text{symmetric}
  \end{array}$};

  \path[every node/.style={font=\sffamily\small}]
    (1) edge [bend left] node [left] {} (2)
    (2) edge [bend left] node [left] {} (1)
    (2) edge[-] [left] node [right] {} (3);
\end{tikzpicture}
};
\draw[->, >=stealth', very thick]{} (a) -- (b);
\end{tikzpicture}

\end{center}
\caption{A diagrammatic representation of a $\mathbb{Z}_6^{\alpha \Omega'}$ orientifold.}
\label{ZkOrientSOSymFig}
\end{figure}
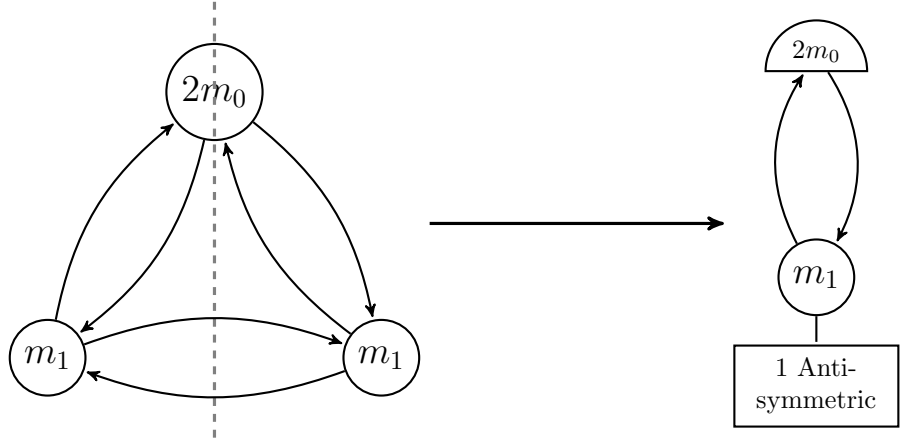

The other choice is
\beq
\tilde{\gamma}_{\alpha \Omega'} = \left(\begin{array}{ccccccc}
    0 & \omega_{2k} \mathds{1}_{m_0} & 0 & 0 & 0 & 0 & 0 \\
    \mathds{1}_{m_0} & 0 & 0 & 0 & 0 & 0 & 0 \\
    0 & 0 & 0 & \omega_{2k}^3 \mathds{1}_{m_1} & 0 & 0 & 0 \\
    0 & 0 & \mathds{1}_{m_1} & 0 & 0 & 0 & 0 \\
    0 & 0 & 0 & 0 & \ddots & 0 & 0 \\
    0 & 0 & 0 & 0 & 0 & 0 & -\mathds{1}_{m_{(k-1)/2}} \\
    0 & 0 & 0 & 0 & 0 & \mathds{1}_{m_{(k-1)/2}} & 0
\end{array} \right)
\eeq
This orientifold results in a linear quiver with $\U(m_i)$ gauge group factors, with an $\USp(2 m_{(k-1)/2})$ factor on one end of the quiver and a symmetric tensor representation on the other end.

\begin{figure}
\begin{center}
\begin{tikzpicture}
\node[] (a) at (0,0) {
\begin{tikzpicture}[->,>=stealth',shorten >=1pt,auto,node distance=5cm,
                    thick,main node/.style={circle,draw,font=\sffamily\Large\bfseries}]

    \node[] (0) {};
    \node[main node] (1) [below of=0, above = 1cm] {$2m_0$};
    \node[main node] (2) [above left of=1, right = 0.8cm] {$m_1$};
    \node[main node] (3) [above right of=1, left = 0.8cm] {$m_1$};

  \draw[dashed, -, very thick, black!50] (0,1.2) -- (0,-4.7);
  \path[every node/.style={font=\sffamily\small}]
    (1) edge [bend left=20] node [left] {} (3)
        edge [bend left=20] node[left] {} (2)
    (2) edge [bend left=20] node [right] {} (1)
        edge [bend left=20] node [right] {} (3)
    (3) edge [bend left=20] node [right] {} (2)
        edge [bend left=20] node [right] {} (1);
\end{tikzpicture}
};
\node[] (b) at (8,0) {
\begin{tikzpicture}[->,>=stealth',shorten >=1pt,auto,node distance=3cm,
                    thick,main node/.style={circle,draw,font=\sffamily\Large\bfseries},square node/.style={rectangle,draw},sp node/.style={shape=semicircle,draw, shape border rotate=180},so node/.style={shape=semicircle,draw}]

  \node[square node] (1) {1 Symmetric};
  \node[main node] (2) [below of=1,above=1cm] {$m_1$};
  \node[sp node] (3) [below of=2] {$2m_0$};

  \path[every node/.style={font=\sffamily\small}]
    (2) edge [bend left] node [left] {} (3)
    (3) edge [bend left] node [left] {} (2)
    (1) edge[-] [left] node [right] {} (2);
\end{tikzpicture}
};
\draw[->, >=stealth', very thick]{} (a) -- (b);
\end{tikzpicture}

\end{center}
\caption{A diagrammatic representation of another $\mathbb{Z}_6^{\alpha \Omega'}$ orientifold.}
\label{ZkOrientSpAntiSymFig}
\end{figure}
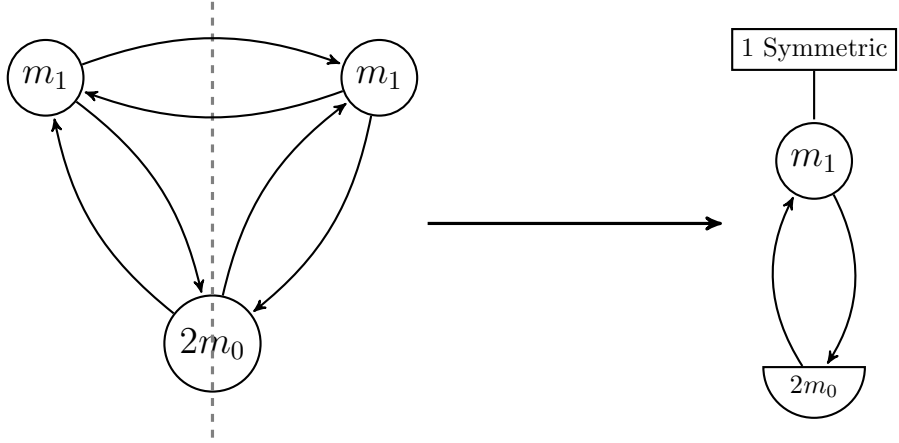

\subsection{$\Dic_k^{j \Omega'} = \left<\alpha, j \Omega'\right>$}

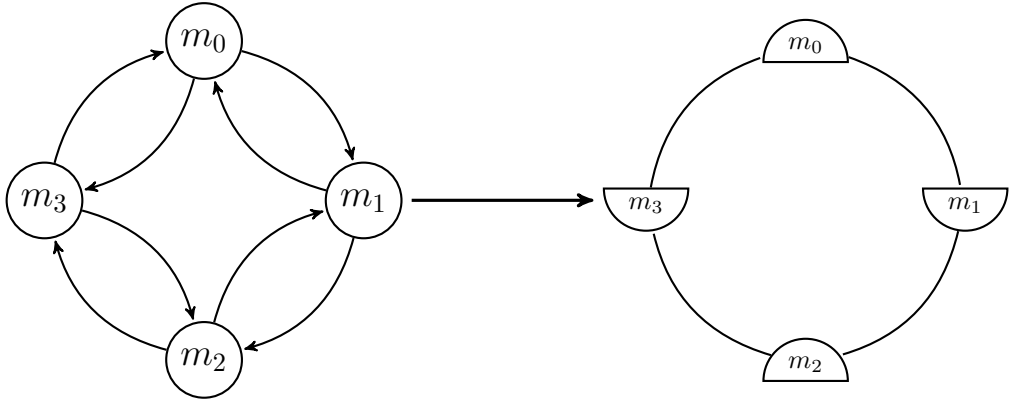
\begin{figure}
\begin{center}
\begin{tikzpicture}
\node[] (a) at (0,0) {
\begin{tikzpicture}[->,>=stealth',shorten >=1pt,auto,node distance=3cm,
                    thick,main node/.style={circle,draw,font=\sffamily\Large\bfseries}]

  \node[main node] (1) {$m_0$};
  \node[main node] (2) [below right of=1] {$m_1$};
  \node[main node] (3) [below left of=2] {$m_2$};
  \node[main node] (4) [below left of=1] {$m_3$};
  
  \path[every node/.style={font=\sffamily\small}]
    (1) edge [bend left] node [left] {} (4)
        edge [bend left] node[left] {} (2)
    (2) edge [bend left] node [right] {} (1)
        edge [bend left] node [right] {} (3)
    (3) edge [bend left] node [right] {} (2)
        edge [bend left] node [right] {} (4)
    (4) edge [bend left] node [left] {} (3)
        edge [bend left] node[left] {} (1);
\end{tikzpicture}
};
\node[] (b) at (8,0) {
\begin{tikzpicture}[->,>=stealth',shorten >=1pt,auto,node distance=3cm,
                    thick,main node/.style={circle,draw,font=\sffamily\Large\bfseries},square node/.style={rectangle,draw},sp node/.style={shape=semicircle,draw, shape border rotate=180},so node/.style={shape=semicircle,draw}]
                    
  \node[so node] (1) {$m_0$};
  \node[sp node] (2) [below right of=1] {$m_1$};
  \node[so node] (3) [below left of=2] {$m_2$};
  \node[sp node] (4) [below left of=1] {$m_3$};
  
  \path[every node/.style={font=\sffamily\small}]
    (1) edge[-] [bend left] node[left] {} (2)
    (2) edge[-] [bend left] node [right] {} (3)
    (3) edge[-] [bend left] node [right] {} (4)
    (4) edge[-] [bend left] node[left] {} (1);
\end{tikzpicture}
};

\draw[->, >=stealth', very thick]{} (a) -- (b);

\end{tikzpicture}
\end{center}
\caption{A diagrammatic representation of a $\Dic_2 = \left<\alpha, j \Omega'\right>$ orientifold. Starting with the orbifold by the subgroup $G_1 = \mathbb{Z}_4$, adding in the generator $j \Omega'$ produces an alternating chain of $\SO$ and $\USp$ gauge group factors. It also identifies the $X$ and $\bar{X}$ bifundamental hypermultiplets, so that we have a single $\textbf{vect}\otimes \textbf{vect}$ hypermultiplet connecting each pair of adjacent factors.}
\label{AlternatingAQuiverFig}
\end{figure}

The dicyclic group $\Dic_k$ contains an index-2 subgroup $\mathbb{Z}_{2k}$. 
Thus, we can define an orientifold by taking the index preserving subgroup to be $\mathbb{Z}_{2k} \subset \Gamma = \Dic_k$.
We denote this orientifold by $\Dic_k = \left<\alpha, j \Omega'\right>$, where $\alpha, j \in \SU(2)$ generate the dicyclic group of order $4k$.
The orientation-preserving subgroup is $G_1=\mathbb{Z}_{2k} = \left<\alpha\right>$, and we can represent $\alpha$ in the standard way, as described above.
The orientation-reversing element can be represented as
\beq
\gamma_{j \Omega'} = \left(\begin{array}{ccccc}
    \mathds{1}_{m_0} & 0 & 0 & \cdots & 0\\
    0 & \Omega_{m_1} & 0 & \cdots & 0 \\
    0 & 0 & \mathds{1}_{m_2} & \cdots & 0 \\
    \vdots & \vdots & \vdots & \ddots & \vdots \\
    0 & 0 & 0 & \cdots & \Omega_{m_{2k-1}}
\end{array} \right).
\eeq
This orientifold gives the affine $A_{2k}$ quiver with alternating $\SO$ and $\USp$ gauge group factors, see Figure \ref{AlternatingAQuiverFig}.

\subsection{$\Dic_{2k}^{\alpha\Omega'} = \left<\alpha \Omega', j \right>$}

The binary dicyclic group $\Dic_{2k}$ contains an index-2 subgroup $\Dic_k$, which we can use to define another orientifold.
The orientifold group is $\Dic_{2k}^{\alpha\Omega'} = \left<\alpha \Omega', j | \alpha^{4k} = 1, j^2 = \alpha^{2k}, \alpha j = j \alpha^{-1} \right>$, and the orientation-preserving subgroup is $G_1 = \Dic_{k} = \langle \alpha^2, j\rangle$.
It is straightforward to find representations of this orientifold group, which lead to the quivers in Table \ref{O3Table}.
See also \cite{Uranga:1999mb}.

\subsection{$\mathbb{Z}_2^{\Omega'} \times \{\text{Tetra, Octa, Icosa}\}$}

We can also consider an orientifold of the form $\mathbb{Z}_2 \times G_1$, where $G_1$ is one of the exceptional subgroups of $\SU(2)$.
Since these orientifolds include an O7-plane, we will need to include D7-branes to cancel their RR charge.
The resulting quivers are summarized in Table \ref{ExceptionalTable}.

\subsection{Tetra $\subset$ Octa}

Of the three exceptional subgroups of $\SU(2)$, only the binary octahedral group contains an index-two subgroup (namely, the binary tetrahedral group).
We can consider the orientifold whose orientation preserving elements are the binary tetrahedral group $\mathbb{B}\mathbb{T} \subset \mathbb{B}\mathbb{O}$, and whose orientation-reversing elements are the other elements of $\mathbb{B}\mathbb{O}$.
There are two different choices for the representation of the orientifold group acting on the Chan-Paton factors, and the resulting quivers are shown in Table \ref{ExceptionalTable}.

\section{Twisted Holographic Dualities}\label{TwistedBackgrounds}

The prototypical example of twisted holography is the duality between the chiral algebra $\mathcal{A}_N$ associated to $\mathcal{N}=4$ super Yang-Mills theory with $\SU(N)$ gauge group and the B-model topological string theory on $\SL_2 \mathbb{C}$ \cite{Costello:2018zrm}.
We wish to extend this proposal to the chiral algebras associated to the class of theories that we referred to above as ``$\mathcal{N}=2$ superconformal gauge theories''.
In general, the twisted holographic duals of these theories are given by topological strings on $\SL_2 \mathbb{C}/\Gamma$ for some discrete subgroup $\Gamma \in \SU(2)$.

It is worth noting that not all of the quivers we've described above give rise to distinct chiral algebras.
This is because of the phenomenon of ``duality'': a given superconformal field theory can have multiple weakly coupled limits.
Since the chiral algebra associated to an $\mathcal{N}=2$ SCFT is independent of the coupling, it may be described using any of the theory's weakly coupled limits, giving rise to different cohomological constructions of the same chiral algebra.

The simplest example of this is the duality between $\mathcal{N}=4$ super Yang-Mills theory with gauge group $\USp(2N)$ and $\mathcal{N}=4$ SYM with gauge group $\SO(2N+1)$.
This duality can be generalized to a large class of theories that arise from webs of D4- and NS5-branes in type IIA string theory, in the presence of a pair of O6-planes and (sometimes) D6-branes \cite{Uranga:1998uj}.
In fact, these ``elliptic'' theories are exactly those which we have been considering with orientifold group $\mathbb{Z}_2^{\Omega'} \times \mathbb{Z}_k$ or $\mathbb{Z}_{2k}^{\alpha \Omega'}$ \cite{Ennes:2000fu, Park:1998zh}.

Indeed, in the infinite-$N$ limit, we conjecture that any quivers corresponding to the same orientifold group (i.e. those appearing in the same boxes in Tables \ref{O3Table}, \ref{O7Table}, and \ref{ExceptionalTable}) give rise to the same chiral algebra.
This is consistent with the fact that they will describe the same B-model background, which is itself determined entirely by the orientifold group.
However, at finite $N$ one must be careful to properly relate the ranks of the gauge groups in a putative dual pair of quivers.
It will only sometimes be possible to choose the ranks in such a way as to obtain a finite $N$ equivalence of chiral algebras.

\subsection{Orientifolds with O7$^-$-Planes: $\Gamma=\mathbb{Z}_2^{\Omega'}\times G_1$}

An unoriented version of the topological string was introduced in \cite{Costello:2019jsy}.
It was shown therein that the theory can be defined and is anomaly free in complex dimensions 3 and 5, provided that an appropriate choice is made for the number of space-filling branes (namely $\SO(8)$ in dimension 3 or $\SO(32)$ in dimension 5).
Classical solutions of this theory, like its oriented counterpart, include a choice of Calabi-Yau manifold.
It is therefore natural to ask what the holographic description of this theory on $\SL_2 \mathbb{C}$ could be.
We propose that the boundary dual description is given by the chiral algebra associated to the 4d $\mathcal{N}=2$ gauge theory with $\USp(2N)$ gauge group and hypermultiplet content consisting of one antisymmetric and four fundamental representations.

As we reviewed above, this theory arises as the low-energy worldvolume theory for a stack of D3-branes in the presence of eight (half-)D7-branes on an O7$^-$-plane, which arises from the orientifold group $\Gamma = \langle \Omega' \rangle$.
As a first piece of evidence for this conjecture, we note that the global symmetry subalgebra of this chiral algebra contains an $\SO(8)$ Kac-Moody algebra, due to the $\SO(8)$ flavor symmetry of the fundamental hypermultiplets.
This corresponds holographically to the $\SO(8)$ holomorphic Chern-Simons theory living on the space-filling branes.

The conjecture can be extended in an obvious way: the chiral algebra associated to the orientifold theory with orientifold group $\Gamma=\mathbb{Z}_2^{\Omega'}\times G_1$ is holographically dual to unoriented topological strings on $\SL_2 \mathbb{C} / G_1$.

\subsection{Orientifolds with O3-Planes}

Now let's consider the twisted holographic backgrounds arising from orientifolds by $\Gamma \subset \SU(2)$, where the orientation-preserving subgroup $G_1$ is an index-2 subgroup $G_1 \subset \Gamma$.
These orientifolds contain an O3-plane, on top of which we place a stack of D3-branes.
As in the prototypical example of twisted holography \cite{Costello:2018zrm}, the location of the D3-branes is not part of the backreacted solution in the bulk.
We find that the resulting background is simply $\SL_2\mathbb{C} / \Gamma$, where $\SU(2)$ acts on $\SL_2\mathbb{C}$ by right multiplication.
How does this differ from the background arising from the \emph{orbifold} by the same subgroup $\Gamma \subset \SU(2)$?

To answer this question, we consider the simplest case. Namely, the orientifold by $\mathbb{Z}^{\alpha \Omega'}_2 = \lbrace 1, \alpha \Omega'\rbrace$.
The relevant field theory is then $\mathcal{N}=4$ super Yang-Mills theory with gauge group $\SO(N)$ or $\USp(2N)$.
The holographic description of this theory was considered in \cite{Witten:1998xy}.
Therein, it was found that the physical holographic duality involves type IIB superstring theory on AdS$_5\times \mathbb{RP}^5$.
Upon twisting, the theory localizes onto a subspace: $\SL_2 \mathbb{C} / \mathbb{Z}_2 = \PSL_2 \mathbb{C}$.
This is the same Calabi-Yau threefold that is relevant to the $\U(N) \times \U(N)$ two-node quiver theory, which arises when considering an \emph{orbifold} by $\mathbb{Z}_2$.
One might be tempted to conclude that the theories have the same chiral algebra.
However, we will see that this is not the case!
Let's examine the chiral algebras associated to each theory, in order to determine what distinguishes them.

We start from the chiral algebra $\mathcal{A}_N$ associated to $\U(N)$ $\mathcal{N}=4$ super Yang-Mills theory.
$\mathcal{A}_N$ can be constructed in terms a free chiral algebra consisting of symplectic bosons $Z^i$ transforming in the $\bf{fund}$ of an $\SU(2)$ flavor symmetry, and a $bc$-system, all of which transform in the adjoint of the $\U(N)$ gauge symmetry.
In the large-$N$ limit, the single-trace primary operators of $\mathcal{A}_N$ form four infinite towers \cite{Costello:2018zrm}:

\beq
\begin{array}{c}
     A_n = \Tr(Z^{(i_1} Z^{i_2} \cdots Z^{i_n)}) \\
     B_n = \Tr(b \ Z^{(i_1} Z^{i_2} \cdots Z^{i_n)}) \\
     C_n = \Tr(\partial c \ Z^{(i_1} Z^{i_2} \cdots Z^{i_n)}) \\
     D_n = \epsilon_{ij} \Tr(Z^i Z^{(j} Z^{i_1} \cdots Z^{i_n)})
\end{array}
\eeq
where parentheses indicate symmetrization of indices.

To find the chiral algebra associated to the two-node quiver theory, we simply impose invariance under $\mathbb{Z}_2$, which acts as $Z^i \rightarrow -Z^i$.
It is clear that the resulting chiral algebra has primary operators consisting of the same four towers listed above, with the condition that $n$ is even.

On the other hand, the chiral algebra for $\mathcal{N}=4$ super Yang-Mills theory with $\SO(N)$ or $\USp(N)$ gauge group is obtained by simply imposing that the symplectic bosons $Z^i$ and $bc$-system lie in the corresponding Lie algebra.
Imposing this condition kills some of the operators, since
\beq
\Tr(M_1 \cdots M_n) = \Tr\left((M_1 \cdots M_n)^{\intercal}\right) = \Tr\left(M_n^{\intercal} \cdots M_1^{\intercal}\right)
\eeq
If $M_i \in \mathfrak{so}(N)$, we have $M_i^{\intercal} = -M_i$, so
\beq
\Tr(M_1 \cdots M_n) = \Tr\left(M_n^{\intercal} \cdots M_1^{\intercal}\right) = (-1)^n \Tr\left(M_n \cdots M_1\right)
\eeq
The same equation holds when $M_i \in \mathfrak{sp}(2N)$, since then $M_i^{\intercal} = - \Omega M_i \Omega^{-1}$.
Applying this relation to the towers of primary operators, we find that the $A_n$ and $D_n$ operators vanish for $n$ odd, while the $B_n$ and $C_n$ operators vanish for $n$ even.
Therefore, we see that the two-node quiver is distinguished from the $\SO(N)$ and $\USp(2N)$ by the $B$ and $C$ towers of operators, while the chiral algebras for the latter two theories are identical in the large-$N$ limit (a manifestation of S-duality, as hinted at above).
Now let's discuss the bulk interpretation of this distinction.

Each of the towers of primary operators has a bulk interpretation in terms of modifications of boundary conditions for the fields of Kodaira-Spencer gravity.
The $A$ and $D$ towers are associated to the Beltrami differential $\beta$ describing deformations of the complex structure of the bulk Calabi-Yau threefold.
The fact that the two-node quiver and $\SO$/$\USp$ chiral algebras have identical $A$ and $D$ towers is then consistent with them describing the same vacuum CY3, namely $\PSL_2 \mathbb{C}$.
The $B$ and $C$ towers correspond to additional degrees of freedom of Kodaira-Spencer theory, which can be formulated as fermionic Chern-Simons fields $\alpha$ and $\gamma$ \cite{Costello:2018zrm}.
It is clear from our discussion above that the distinction between the two-node orbifold theory and the orientifold theories is precisely in the boundary conditions for these Chern-Simons fields around the nontrivial cycle of $\PSL_2 \mathbb{C}$.

Indeed, in the general case we see that the choice of index-2 subgroup of $G_1 \subset \Gamma$ involved in defining the orientifold is precisely a choice of nontrivial $\mathbb{Z}_2$-bundle over $\SL_2 \mathbb{C} / \Gamma$, where $G_1$ are taken to be the nontrivial cycles with trivial $\mathbb{Z}_2$ homotopy.
This $\mathbb{Z}_2$-bundle then defines a choice of boundary conditions for the fermionic fields $\alpha$ and $\gamma$.
More precisely, the orientifold defines a choice of spin structure over $\SL_2 \mathbb{C} / \Gamma$.

\section{Discussion}

In this note, we reviewed orientifold constructions of a large class of four-dimensional $\mathcal{N}=2$ superconformal field theories, with the aim of understanding the twisted holographic dualities involving their associated chiral algebras.
The theories we described fell into two classes depending on whether the orientifold group contains the element $\Omega' = \Omega R_{45} (-1)^{F_L}$, which swaps the orientation of the worldsheet while acting trivially on $\mathbb{C}^2$.
When it does, the theory is associated with a choice of $\Gamma$, a finite subgroup of $\SU(2)$, and the orientifold group is then $\Gamma \oplus \Omega' \ \Gamma$.
When the orientifold group does not contain $\Omega'$, we found that the theories are associated with pairs $(\Gamma, G_1)$, where $\Gamma$ is a finite subgroup of $\SU(2)$, and $G_1$ is an index-two subgroup of $\Gamma$ that plays the role of the orientation-preserving elements of the orientifold group.

The corresponding chiral algebras admit a cohomological construction starting from a system of symplectic bosons.
We argued that the bulk dual of these chiral algebras is related to topological string theory on $\SL(2,\mathbb{C})/\Gamma$.
We emphasize that our purpose herein was simply to conjecture these holographic dualities, and we leave the proof of the dualities to future work.

The orientifold setups we have described herein offer a fertile testing ground for explicit holographic calculations.
There many interesting future directions, including:

\begin{itemize}
\item It would be interesting to calculate chiral algebra observables involving very ``heavy'' operators (i.e. those with large conformal dimension). In the prototypical example of twisted holography, this was explored for operators with $\Delta \sim O(N)$ in \cite{Budzik:2021fyh}, where the operators were found to source branes that wrap holomorphic surfaces in the bulk $\SL_2 \mathbb{C}$. It would be interesting to extend this to the chiral algebras dual to orientifolds we have described.
\item Taking the operators to have even larger scaling dimension $\Delta \sim O(N^2)$ should result in nontrivial backreaction in the bulk. What are the Calabi-Yau threefolds that are dual to correlation functions of such operators?
\item The examples of twisted holography we have described so far have involved field theories that admit weakly coupled limits. This makes them particularly tractible examples, since the chiral algebras can be constructed using cohomology of a free chiral algebra. It would be very interesting to understand twisted holography for theories that do not admit such a construction. The most immediate target are theories of class $\mathcal{S}$, for which some details of the chiral algebras are known (see e.g. \cite{Beem:2014rza, Arakawa:2018egx}). The main difficulty in approaching these theories holographically is making sense of the large-N limit.
\end{itemize}

\subsection*{Acknowledgements}
I would like to thank Davide Gaiotto for  suggesting this as an interesting direction to pursue. I would also like to thank Adri\'an Lopez Raven for many helpful discussions and collaboration during the initial stages of the project.
I would like to thank Roland Bittleston, Kevin Costello, Nikita Grygoryev, Alexendre Homrich, Seraphim Jarov, Justin Kulp, and Pedro Vieira for enlightening conversations.
Much of the work leading to this note was completed while the author was a student at the University of Waterloo and Perimeter Institute for Theoretical Physics.
Research at Perimeter Institute is supported in part by the Government of Canada through the Department of Innovation, Science, and Economic Development Canada and by the Province of Ontario through the Ministry of Colleges and Universities.

\newpage
\renewcommand{\arraystretch}{2}

\begin{table}
\centering
    \begin{tabular}{|| >{\centering\Large} m{5cm} | >{\centering\arraybackslash} m{7cm} ||}
         \hline
         Orbifold Group & \Large{Quiver Diagram} \\
         \hline\hline
         $\mathbb{Z}_k$ & \begin{tikzpicture}[baseline=0,->,>=stealth',shorten >=1pt,auto,node distance=3cm,
                    thick,main node/.style={circle,draw,font=\sffamily\Large\bfseries},empty node/.style={font=\sffamily\Large\bfseries}]

  \node[main node] (1) {$N$};
  \node[main node] (2) [below right of=1] {$N$};
  \node[empty node] (3) [below left of=2] {\dots};
  \node[empty node] (3L) [below left of=2, left = 0.3 cm] {};
  \node[empty node] (3R) [below left of=2, right = 0.3 cm] {};
  \node[main node] (4) [below left of=1] {$N$};
  
  \path[every node/.style={font=\sffamily\small}]
    (1) edge [bend left] node [left] {} (4)
        edge [bend left] node[left] {} (2)
    (2) edge [bend left] node [right] {} (1)
        edge [bend left] node [right] {} (3R)
    (3R) edge [bend left] node [right] {} (2)
    (3L) edge [bend left] node [right] {} (4)
    (4) edge [bend left] node [left] {} (3L)
        edge [bend left] node[left] {} (1);

    \addvmargin{2mm}
\end{tikzpicture} \\
         \hline
         $\Dic_k$ & \begin{tikzpicture}[baseline=0,->,>=stealth',shorten >=1pt,auto,node distance=2cm,
                    thick,main node/.style={circle,draw,font=\sffamily\Large\bfseries},empty node/.style={font=\sffamily\Large\bfseries}]

  \node[main node] (1) {$N$};
  \node[main node] (2) [below right of=1] {$2N$};
  \node[main node] (3) [below left of=2] {$N$};
  \node[main node] (7) [right of=2] {$2N$};
  \node[empty node] (8) [right of=7] {\dots};
  \node[main node] (4) [right of=8]{$2N$};
  \node[main node] (5) [above right of=4]{$N$};
  \node[main node] (6) [below right of=4]{$N$};

  \path[every node/.style={font=\sffamily\small}]
    (1) edge [bend left=20] node [left] {} (2)
    (2) edge [bend left=20] node [right] {} (1)
        edge [bend left=20] node [right] {} (3)
        edge [bend left=20] node [right] {} (7)
    (3) edge [bend left=20] node [right] {} (2)
    (7) edge [bend left=20] node [right] {} (8)
        edge [bend left=20] node [right] {} (2)
    (8) edge [bend left=20] node [right] {} (4)
        edge [bend left=20] node [right] {} (7)
    (4) edge [bend left=20] node [right] {} (8)
        edge [bend left=20] node [right] {} (5)
        edge [bend left=20] node [right] {} (6)
    (5) edge [bend left=20] node [right] {} (4)
    (6) edge [bend left=20] node [right] {} (4);

    \addvmargin{2mm}
\end{tikzpicture} \\
         \hline
         Binary Tetrahedral & \begin{tikzpicture}[baseline=0,->,>=stealth',shorten >=1pt,auto,node distance=2cm,
                    thick,main node/.style={circle,draw,font=\sffamily\Large\bfseries}]

  \node[main node] (1) {$3N$};
  \node[main node] (2a) [above of=1, above = -1cm] {$2N$};
  \node[main node] (2b) [below left of=1, above = 0.15 cm] {$2N$};
  \node[main node] (2c) [below right of=1, above = 0.15 cm] {$2N$};
  \node[main node] (3a) [above of=2a, above = -0.8cm] {$N$};
  \node[main node] (3b) [below left of=2b, above = 0.15 cm] {$N$};
  \node[main node] (3c) [below right of=2c, above = 0.15 cm] {$N$};

  \path[every node/.style={font=\sffamily\small}]
    (1) edge [bend left=20] node [left] {} (2a)
        edge [bend left=20] node [left] {} (2b)
        edge [bend left=20] node [left] {} (2c)
    (2a) edge [bend left=20] node [right] {} (1)
        edge [bend left=20] node [left] {} (3a) 
    (2b) edge [bend left=20] node [right] {} (1)
        edge [bend left=20] node [left] {} (3b)
    
    (2c) edge [bend left=20] node [right] {} (1)
        edge [bend left=20] node [left] {} (3c)
    (3a) edge [bend left=20] node [right] {} (2a)
    (3b) edge [bend left=20] node [right] {} (2b)
    (3c) edge [bend left=20] node [right] {} (2c);
    
    \addvmargin{2mm}
\end{tikzpicture} \\
         \hline
         Binary Octahedral & \begin{tikzpicture}[baseline=0,->,>=stealth',shorten >=1pt,auto,node distance=1.7cm,
                    thick,main node/.style={circle,draw,font=\sffamily\Large\bfseries}]

  \node[main node] (1) {$4N$};
  \node[main node] (2a) [above of=1, above = 0cm] {$2N$};
  \node[main node] (2b) [left of=1] {$3N$};
  \node[main node] (2c) [right of=1] {$3N$};
  \node[main node] (3b) [left of=2b] {$2N$};
  \node[main node] (3c) [right of=2c] {$2N$};
  \node[main node] (4b) [left of=3b] {$N$};
  \node[main node] (4c) [right of=3c] {$N$};

  \path[every node/.style={font=\sffamily\small}]
    (1) edge [bend left=20] node [left] {} (2a)
        edge [bend left=20] node [left] {} (2b)
        edge [bend left=20] node [left] {} (2c)
    (2a) edge [bend left=20] node [right] {} (1)
    (2b) edge [bend left=20] node [right] {} (1)
        edge [bend left=20] node [left] {} (3b)
    
    (2c) edge [bend left=20] node [right] {} (1)
        edge [bend left=20] node [left] {} (3c)
    (3b) edge [bend left=20] node [right] {} (2b)
         edge [bend left=20] node [left] {} (4b)
    (3c) edge [bend left=20] node [right] {} (2c)
         edge [bend left=20] node [left] {} (4c)
    (4b) edge [bend left=20] node [right] {} (3b)
    (4c) edge [bend left=20] node [right] {} (3c);

    \addvmargin{2mm}
\end{tikzpicture} \\
         \hline
         Binary Icosahedral & \begin{tikzpicture}[baseline=0,->,>=stealth',shorten >=1pt,auto,node distance=1.7cm,
                    thick,main node/.style={circle,draw,font=\sffamily\Large\bfseries}]

  \node[main node] (1) {$6N$};
  \node[main node] (2a) [above of=1, above = 0cm] {$3N$};
  \node[main node] (2b) [left of=1] {$4N$};
  \node[main node] (2c) [right of=1] {$5N$};
  \node[main node] (3b) [left of=2b] {$2N$};
  \node[main node] (3c) [right of=2c] {$4N$};
  \node[main node] (4c) [right of=3c] {$3N$};
  \node[main node] (5c) [right of=4c] {$2N$};
  \node[main node] (6c) [right of=5c] {$N$};

  \path[every node/.style={font=\sffamily\small}]
    (1) edge [bend left=20] node [left] {} (2a)
        edge [bend left=20] node [left] {} (2b)
        edge [bend left=20] node [left] {} (2c)
    (2a) edge [bend left=20] node [right] {} (1)
    (2b) edge [bend left=20] node [right] {} (1)
        edge [bend left=20] node [left] {} (3b)
    (2c) edge [bend left=20] node [right] {} (1)
        edge [bend left=20] node [left] {} (3c)
    (3b) edge [bend left=20] node [right] {} (2b)
    (3c) edge [bend left=20] node [right] {} (2c)
        edge [bend left=20] node [left] {} (4c)
    (4c) edge [bend left=20] node [right] {} (3c)
        edge [bend left=20] node [left] {} (5c)
    (5c) edge [bend left=20] node [right] {} (4c)
        edge [bend left=20] node [left] {} (6c)
    (6c) edge [bend left=20] node [right] {} (5c);

    \addvmargin{2mm}
\end{tikzpicture} \\
         \hline
    \end{tabular}
\caption{Summary of orbifolds and their corresponding quivers. Here we set the size of each gauge group factor to the appropriate value to ensure conformal symmetry.}
\label{OrbifoldTable}
\end{table}

\begin{table}
\centering
    \begin{tabular}{|| >{\centering\Large} m{4cm} | >{\centering\Large} m{3cm} | >{\centering\arraybackslash} m{7cm} ||}
         \hline
         Orientifold Group & $k$ Even/Odd? & \Large{Quiver Diagram(s)} \\
         \hline\hline
         \multirow{5}{4cm}{\centering{$\mathbb{Z}_{2k}^{\alpha\Omega'}$}} & \multirow{2}{3cm}{\centering{Even}} & \begin{tikzpicture}[baseline=0,->,>=stealth',shorten >=1pt,auto,node distance=3cm,
                    thick,main node/.style={circle,draw,font=\sffamily\Large\bfseries},square node/.style={rectangle,draw},sp node/.style={shape=semicircle,draw, shape border rotate=180},so node/.style={shape=semicircle,draw},
                    empty node/.style={font=\sffamily\Large\bfseries}]

  \node[sp node, scale=1.3] (1) {$2N$};
  \node[main node] (2) [right of=1] {$2N+2$};
  \node[main node] (3) [right of=2] {$2N+4$};
  \node[empty node] (4) [right of=3, left=0.3cm] {\dots};
  \node[main node,scale =0.7] (5) [right of=4] {$2N+k-2$};
  \node[so node] (6) [right of=5] {$2N+k$};

  \path[every node/.style={font=\sffamily\small}]
    (1) edge [bend left=20] node [left] {} (2)
    (2) edge [bend left=20] node [right] {} (3)
    (2) edge [bend left=20] node [left] {} (1)
    (3) edge [bend left=20] node [right] {} (2)
    (3) edge [bend left=20] node [right] {} (4)
    (4) edge [bend left=20] node [right] {} (3)
    (4) edge [bend left=20] node [right] {} (5)
    (5) edge [bend left=20] node [right] {} (4)
    (5) edge [bend left=20] node [right] {} (6)
    (6) edge [bend left=20] node [right] {} (5);

    \addvmargin{2mm}
\end{tikzpicture} \\
         & & \normalsize{\begin{tikzpicture}[->,>=stealth',shorten >=1pt,auto,node distance=3cm,
                    thick,main node/.style={circle,draw,font=\sffamily\Large\bfseries},square node/.style={rectangle,draw},sp node/.style={shape=semicircle,draw, shape border rotate=180},so node/.style={shape=semicircle,draw},
                    empty node/.style={font=\sffamily\Large\bfseries}]

  \node[square node] (1) {$1$ \textbf{Asymm}};
  \node[main node] (2) [right of=1,left=0cm] {$N+2$};
  \node[main node] (3) [right of=2] {$N+4$};
  \node[empty node] (4) [right of=3,left=0.2cm] {\dots};
  \node[main node] (5) [right of=4,left=0] {$N+k$};
  \node[square node] (6) [right of=5,left = 0cm] {$1$ \textbf{Symm}};

  \path[every node/.style={font=\sffamily\small}]
    (2) edge[-] node [left] {} (1)
    (2) edge [bend left=20] node [right] {} (3)
    (3) edge [bend left=20] node [right] {} (2)
    (3) edge [bend left=20] node [right] {} (4)
    (4) edge [bend left=20] node [right] {} (3)
    (4) edge [bend left=20] node [right] {} (5)
    (5) edge [bend left=20] node [right] {} (4)
    (5) edge[-] node [right] {} (6);
\end{tikzpicture}} \\
         \cline{2-3}
         & \multirow{2}{3cm}{\centering{Odd}} & \normalsize{\begin{tikzpicture}[baseline=0,->,>=stealth',shorten >=1pt,auto,node distance=3cm,
                    thick,main node/.style={circle,draw,font=\sffamily\Large\bfseries},square node/.style={rectangle,draw},sp node/.style={shape=semicircle,draw, shape border rotate=180},so node/.style={shape=semicircle,draw},
                    empty node/.style={font=\sffamily\Large\bfseries}]
  
  \node[sp node, scale=1.3] (1) {$2N$};
  \node[main node] (2) [right of=1] {$2N+2$};
  \node[main node] (3) [right of=2] {$2N+4$};
  \node[empty node] (4) [right of=3, left=0.3cm] {\dots};
  \node[main node,scale =0.7] (5) [right of=4] {$2N+k-1$};
  \node[square node] (6) [right of=5,left=0cm] {$1$ \textbf{Symm}};

  \path[every node/.style={font=\sffamily\small}]
    (1) edge [bend left=20] node [left] {} (2)
    (2) edge [bend left=20] node [right] {} (3)
    (2) edge [bend left=20] node [left] {} (1)
    (3) edge [bend left=20] node [right] {} (2)
    (3) edge [bend left=20] node [right] {} (4)
    (4) edge [bend left=20] node [right] {} (3)
    (4) edge [bend left=20] node [right] {} (5)
    (5) edge [bend left=20] node [right] {} (4)
    (5) edge[-] node [left] {} (6);

    \addvmargin{2mm}
\end{tikzpicture}} \\
         & & \normalsize{\begin{tikzpicture}[->,>=stealth',shorten >=1pt,auto,node distance=3cm,
                    thick,main node/.style={circle,draw,font=\sffamily\Large\bfseries},square node/.style={rectangle,draw},sp node/.style={shape=semicircle,draw, shape border rotate=180},so node/.style={shape=semicircle,draw},
                    empty node/.style={font=\sffamily\Large\bfseries}]

  \node[square node] (1) {$1$ \textbf{Asymm}};
  \node[main node, inner sep=0.38cm] (2) [right of=1,left=0] {$N$};
  \node[main node] (3) [right of=2] {$N+2$};
  \node[empty node] (4) [right of=3, left=0.3cm] {\dots};
  \node[main node,scale=0.67] (5) [right of=4] {$N+k-3$};
  \node[so node] (6) [right of=5] {$N+k-1$};

  \path[every node/.style={font=\sffamily\small}]
    (2) edge[-] node [left] {} (1)
    (2) edge [bend left=20] node [right] {} (3)
    (3) edge [bend left=20] node [right] {} (2)
    (3) edge [bend left=20] node [right] {} (4)
    (4) edge [bend left=20] node [right] {} (3)
    (4) edge [bend left=20] node [right] {} (5)
    (5) edge [bend left=20] node [right] {} (4)
    (5) edge [bend left=20] node [right] {} (6)
    (6) edge [bend left=20] node [right] {} (5);
\end{tikzpicture}} \\
         \hline
         $\Dic_k^{ j\Omega'}$ & Either & \normalsize{\begin{tikzpicture}[baseline=0,->,>=stealth',shorten >=1pt,auto,node distance=3cm,
                    thick,main node/.style={circle,draw,font=\sffamily\Large\bfseries},square node/.style={rectangle,draw},sp node/.style={shape=semicircle,draw, shape border rotate=180},so node/.style={shape=semicircle,draw}, empty node/.style={font=\sffamily\Large\bfseries}]
                    
  \node[so node,inner sep=0.22cm] (1) {$2N$};
  \node[sp node] (2) [below right of=1, above=0.5cm] {$2N-2$};
  \node[so node,inner sep=0.22cm] (3) [below of=2, above=0.5cm] {$2N$};
  \node[sp node] (4) [below left of=1, above=0.5cm] {$2N-2$};
  \node[so node,inner sep=0.22cm] (5) [below of=4, above=0.5cm] {$2N$};
  \node[empty node] (6) [below left of = 3, above=0.6cm] {\dots};
  
  \path[every node/.style={font=\sffamily\small}]
    (1) edge[-] [bend left=20] node[left] {} (2)
    (2) edge[-] [bend left=15] node [right] {} (3)
    (4) edge[-] [bend left=20] node[left] {} (1)
    (5) edge[-] [bend left=15] node [left] {} (4)
    (3) edge[-] [bend left=20] node [left] {} (6)
    (6) edge[-] [bend left=20] node [left] {} (5);

    \addvmargin{2mm}
\end{tikzpicture}} \\
         \hline
         \multirow{4}{4cm}{\centering{$\Dic_{2k}^{\alpha \Omega'}$}} & \multirow{2}{3cm}{\centering{Odd}} & \normalsize{\begin{tikzpicture}[baseline=0,->,>=stealth',shorten >=1pt,auto,node distance=1.5cm,
                    thick,main node/.style={circle,draw,font=\sffamily\Large\bfseries},sp node/.style={shape=semicircle,draw, shape border rotate=180},so node/.style={shape=semicircle,draw},empty node/.style={font=\sffamily\Large\bfseries},square node/.style={rectangle,draw}]

  \node[so node,scale = 0.8] (2) [] {$2N+4$};
  \node[sp node] (1) [above left of=2] {$N$};
  \node[sp node] (3) [below left of=2] {$N$};
  \node[sp node] (7) [right of=2,scale = 0.8] {$2N+4$};
  \node[empty node] (8) [right of = 7] {\dots};
  \node[so node] (9) [right of=8,scale=0.6,right=-0.7cm] {$2N+2(k-1)$};
  \node[sp node, scale=0.6] (4) [right of=9, right=0.1cm]{$2N+2(k-1)$};
  \node[so node] (5) [above right of=4,scale=0.6,above=-0.35cm]{$N+k+1$};
  \node[so node] (6) [below right of=4,scale=0.6,above=-0.3cm]{$N+k+1$};

  \path[every node/.style={font=\sffamily\small}]
    (1) edge[-] node [left] {} (2)
    (2) edge[-] node [right] {} (7)
    (3) edge[-] node [left] {} (2)
    (7) edge[-] node [right] {} (8)
    (8) edge[-] node [right] {} (9)
    (9) edge[-] node [right] {} (4)
    (4) edge[-] node [right] {} (5)
        edge[-] node [right] {} (6);

    \addvmargin{2mm}
\end{tikzpicture}} \\
         & & \normalsize{\begin{tikzpicture}[->,>=stealth',shorten >=1pt,auto,node distance=1.8cm,
                    thick,main node/.style={circle,draw,font=\sffamily\Large\bfseries},sp node/.style={shape=semicircle,draw, shape border rotate=180},so node/.style={shape=semicircle,draw},empty node/.style={font=\sffamily\Large\bfseries},square node/.style={rectangle,draw}]

  \node[main node,inner sep=0.1cm] (1) {$N$};
  \node[sp node,inner sep=0.1cm] (2) [right of=1, left=-0.1cm] {$2 N$};
  \node[so node, inner sep=0.1cm] (7) [right of=2, left=-0.5cm] {$2 N+4$};
  \node [empty node] (8) [right of=7,right=-0.9cm] {\dots};
  \node [sp node,scale=0.6] (9) [right of=8,right=-0.3cm] {$2 N+2(k-3)$};
  \node[so node,scale=0.6] (10) [right of=9,right=0cm]{$2N+2(k-1)$};
  \node[main node] (11) [right of=10,scale=0.6] {$N+k-1$};
  
  \path[every node/.style={font=\sffamily\small}]
    (1) edge [bend left=20] node [left] {} (2)
    (2) edge[-] node [right] {} (7)
        edge [bend left=20] node [left] {} (1)
    (7) edge[-] node [right] {} (8)
    (8) edge [-] node [right] {} (9)
    (9) edge [-] node [right] {} (10)
    (10) edge [bend left=20] node [right] {} (11)
    (11) edge [bend left=20] node [right] {} (10);
\end{tikzpicture}} \\
         \cline{2-3}
         & \multirow{2}{3cm}{\centering{Even}} & \normalsize{\begin{tikzpicture}[baseline=0,->,>=stealth',shorten >=1pt,auto,node distance=1.5cm,
                    thick,main node/.style={circle,draw,font=\sffamily\Large\bfseries},sp node/.style={shape=semicircle,draw, shape border rotate=180},so node/.style={shape=semicircle,draw},empty node/.style={font=\sffamily\Large\bfseries},square node/.style={rectangle,draw}]

  \node[so node,inner sep=0.04cm] (2) [] {$2N+4$};
  \node[sp node] (1) [above left of=2] {$N$};
  \node[sp node] (3) [below left of=2] {$N$};
  \node[sp node] (7) [right of=2,scale = 0.8, right=-0.6cm] {$2N+4$};
  \node[empty node] (8) [right of = 7, right =-0.6cm] {\dots};
  \node [sp node,scale=0.5] (9) [right of=8, right=0.04cm] {$2 N+2(k-2)$};
  \node[so node] (10) [right of=9,scale=0.6,right=-0.6cm] {$2N+2k$};
  \node[main node] (11) [right of=10,scale=0.6, right=-0.5cm] {$N+k$};

  \path[every node/.style={font=\sffamily\small}]
    (1) edge[-] node [left] {} (2)
    (2) edge[-] node [right] {} (7)
    (3) edge[-] node [left] {} (2)
    (7) edge[-] node [right] {} (8)
    (8) edge[-] node [right] {} (9)
    (9) edge [-] node [right] {} (10)
    (10) edge [bend left=20] node [right] {} (11)
    (11) edge [bend left=20] node [right] {} (10);

    \addvmargin{2mm}
\end{tikzpicture}} \\
         & & \normalsize{\begin{tikzpicture}[->,>=stealth',shorten >=1pt,auto,node distance=1.5cm,
                    thick,main node/.style={circle,draw,font=\sffamily\Large\bfseries},sp node/.style={shape=semicircle,draw, shape border rotate=180},so node/.style={shape=semicircle,draw},empty node/.style={font=\sffamily\Large\bfseries},square node/.style={rectangle,draw}]

  \node[main node, inner sep=0.1cm](1)[]{$N$};
  \node[sp node,inner sep=0.1cm] (2) [right of=1] {$2N$};
  \node[so node, inner sep=0.1cm] (7) [right of=2,scale = 0.8, right=-0.8cm] {$2N+4$};
  \node[empty node] (8) [right of = 7, right=-0.6cm] {\dots};
  \node[so node] (9) [right of=8,scale=0.6,right=-0.7cm] {$2N+2(k-2)$};
  \node[sp node, scale=0.6] (4) [right of=9, right=0.1cm]{$2N+2(k-2)$};
  \node[so node] (5) [above right of=4,scale=0.6,above=-0.35cm]{$N+k$};
  \node[so node] (6) [below right of=4,scale=0.6,above=-0.3cm]{$N+k$};

  \path[every node/.style={font=\sffamily\small}]
    (1) edge[bend left=20] node [left] {} (2)
    (2) edge[-] node [right] {} (7)
        edge[bend left=20] node [left] {} (1)
    (7) edge[-] node [right] {} (8)
    (8) edge[-] node [right] {} (9)
    (9) edge[-] node [right] {} (4)
    (4) edge[-] node [right] {} (5)
        edge[-] node [right] {} (6);
\end{tikzpicture}} \\
         \hline
    \end{tabular}
\caption{Summary of orientifolds that contain an O3-plane, and their corresponding quivers. These are the orientifolds that don't require the introduction of D7-branes, as is evident from the lack of hypermultiplets transforming in the \textbf{fundamental} representation of any of the gauge group factors. Here we set the size of each gauge group factor to the appropriate value to ensure conformal symmetry. \textbf{Asymm}: \textbf{Anti-symmetric} tensor representation. \textbf{Symm}: \textbf{Symmetric} tensor representation.}
\label{O3Table}
\end{table}

\begin{table}
\centering
    \begin{tabular}{|| >{\centering\Large} m{2.5cm} | >{\centering\Large\arraybackslash} m{3cm} | >{\centering\arraybackslash} m{8.5cm} ||}
         \hline
         Orientifold Group & $k$ Even/Odd? & \Large{Quiver Diagram(s)} \\
         \hline\hline
         \multirow[b]{3}{2.5cm}{\centering{$\mathbb{Z}_{2}^{\Omega'} \times \mathbb{Z}_k$}} & \multirow[b]{2}{3cm}{\centering{Even}} & \begin{tikzpicture}[baseline=0,->,>=stealth',shorten >=1pt,auto,node distance=2.0cm,
                    thick,main node/.style={circle,draw,font=\sffamily\Large\bfseries},square node/.style={rectangle,draw},sp node/.style={shape=semicircle,draw, shape border rotate=180},so node/.style={shape=semicircle,draw},
                    empty node/.style={font=\sffamily\Large\bfseries}]

  \node[sp node, scale=1.3] (1) {$2N$};
  \node[square node,inner sep=0.2cm] (1f) [above of = 1,above = -1.3cm] {2 \textbf{fund}};
  \node[main node] (2) [right of=1] {$2N$};
  \node[main node] (3) [right of=2] {$2N$};
  \node[empty node] (4) [right of=3, left=0cm] {\dots};
  \node[main node] (5) [right of=4,left=0cm] {$2N$};
  \node[sp node, scale=1.3] (6) [right of=5, left=0] {$2N$};
  \node[square node,inner sep=0.2cm] (6f) [above of = 6,above = -1.3cm] {2 \textbf{fund}};

  \path[every node/.style={font=\sffamily\small}]
    (1) edge [bend left=20] node [left] {} (2)
    (2) edge [bend left=20] node [right] {} (3)
    (2) edge [bend left=20] node [left] {} (1)
    (3) edge [bend left=20] node [right] {} (2)
    (3) edge [bend left=20] node [right] {} (4)
    (4) edge [bend left=20] node [right] {} (3)
    (4) edge [bend left=20] node [right] {} (5)
    (5) edge [bend left=20] node [right] {} (4)
    (5) edge [bend left=20] node [right] {} (6)
    (6) edge [bend left=20] node [right] {} (5)
    (1) edge[-] node [right] {} (1f)
    (6) edge[-] node [right] {} (6f);

    \addvmargin{2mm}
\end{tikzpicture} \\
         & & \begin{tikzpicture}[->,>=stealth',shorten >=1pt,auto,node distance=2cm,
                    thick,main node/.style={circle,draw,font=\sffamily\Large\bfseries},square node/.style={rectangle,draw},sp node/.style={shape=semicircle,draw, shape border rotate=180},so node/.style={shape=semicircle,draw},
                    empty node/.style={font=\sffamily\Large\bfseries}]

  \node[square node] (1) {$1$ \textbf{Asymm}};
  \node[main node] (2) [right of=1,left=-0.2cm] {$N$};
  \node[square node,inner sep=0.2cm] (2f) [above of = 2,above = -1.3cm] {2 \textbf{fund}};
  \node[main node] (3) [right of=2] {$N$};
  \node[empty node] (4) [right of=3,left=0.2cm] {\dots};
  \node[main node] (5) [right of=4,left=0.2] {$N$};
  \node[square node,inner sep=0.2cm] (5f) [above of = 5,above = -1.3cm] {2 \textbf{fund}};
  \node[square node] (6) [right of=5,left=-0.7cm] {$1$ \textbf{Asymm}};

  \path[every node/.style={font=\sffamily\small}]
    (1) edge[-] node [left] {} (2)
    (2) edge[-] node [right] {} (2f)
    (2) edge [bend left=20] node [right] {} (3)
    (3) edge [bend left=20] node [right] {} (2)
    (3) edge [bend left=20] node [right] {} (4)
    (4) edge [bend left=20] node [right] {} (3)
    (4) edge [bend left=20] node [right] {} (5)
    (5) edge [bend left=20] node [right] {} (4)
    (5) edge[-] node [right] {} (5f)
    (5) edge[-] node [right] {} (6);
\end{tikzpicture} \\
         \cline{2-3}
         & Odd & \begin{tikzpicture}[baseline=0,->,>=stealth',shorten >=1pt,auto,node distance=2cm,
                    thick,main node/.style={circle,draw,font=\sffamily\Large\bfseries},square node/.style={rectangle,draw},sp node/.style={shape=semicircle,draw, shape border rotate=180},so node/.style={shape=semicircle,draw},
                    empty node/.style={font=\sffamily\Large\bfseries}]

  \node[sp node, scale=1.3] (1) {$2N$};
  \node[square node,inner sep=0.2cm] (1f) [above of = 1,above = -1.3cm] {$2$ \textbf{fund}};
  \node[main node] (2) [right of=1] {$2N$};
  \node[main node] (3) [right of=2] {$2N$};
  \node[empty node] (4) [right of=3, left=0cm] {\dots};
  \node[main node] (5) [right of=4,left=0cm] {$2N$};
  \node[square node,inner sep=0.2cm] (5f) [above of = 5,above = -1.3cm] {$2$ \textbf{fund}};
  \node[square node] (5a) [right of=5, left=-0.7cm] {$1$ \textbf{Asymm}};

  \path[every node/.style={font=\sffamily\small}]
    (1) edge [bend left=20] node [left] {} (2)
    (2) edge [bend left=20] node [right] {} (3)
    (2) edge [bend left=20] node [left] {} (1)
    (3) edge [bend left=20] node [right] {} (2)
    (3) edge [bend left=20] node [right] {} (4)
    (4) edge [bend left=20] node [right] {} (3)
    (4) edge [bend left=20] node [right] {} (5)
    (5) edge [bend left=20] node [right] {} (4)
    (1) edge[-] node [right] {} (1f)
    (5) edge[-] node [right] {} (5f)
    (5) edge[-] node [right] {} (5a);

    \addvmargin{2mm}
\end{tikzpicture} \\
         \hline
         \multirow[b]{5}{2.5cm}{\centering{$\mathbb{Z}_2^{\Omega'} \times \Dic_k$}} & \multirow[b]{3}{3cm}{\centering{Even}} & \begin{tikzpicture}[baseline=0,->,>=stealth',shorten >=1pt,auto,node distance=1.5cm,
                    thick,main node/.style={circle,draw,font=\sffamily\Large\bfseries},sp node/.style={shape=semicircle,draw, shape border rotate=180},so node/.style={shape=semicircle,draw},empty node/.style={font=\sffamily\Large\bfseries},square node/.style={rectangle,draw}]

  \node[so node,scale = 0.8] (2) [] {$2N+2$};
  \node[sp node] (1) [above left of=2] {$N$};
  \node[square node,inner sep=0.2cm] (1f) [left of = 1] {$1$ \textbf{fund}};
  \node[sp node] (3) [below left of=2] {$N$};
  \node[square node,inner sep=0.2cm] (3f) [left of = 3] {$1$ \textbf{fund}};
  \node[sp node] (7) [right of=2] {$2N$};
  \node[empty node] (8) [right of = 7] {\dots};
  \node[sp node] (9) [right of=8] {$2N$};
  \node[so node, scale=0.8] (4) [right of=9, right=-0.3cm]{$2N+2$};
  \node[sp node] (5) [above right of=4]{$N$};
  \node[square node,inner sep=0.2cm] (5f) [right of = 5] {$1$ \textbf{fund}};
  \node[sp node] (6) [below right of=4]{$N$};
  \node[square node,inner sep=0.2cm] (6f) [right of = 6] {$1$ \textbf{fund}};

  \path[every node/.style={font=\sffamily\small}]
    (1) edge[-] node [left] {} (2)
    (2) edge[-] node [right] {} (7)
    (3) edge[-] node [left] {} (2)
    (7) edge[-] node [right] {} (8)
    (8) edge[-] node [right] {} (9)
    (9) edge[-] node [right] {} (4)
    (4) edge[-] node [right] {} (5)
        edge[-] node [right] {} (6)
    (1) edge[-] node [right] {} (1f)
    (3) edge[-] node [right] {} (3f)
    (5) edge[-] node [right] {} (5f)
    (6) edge[-] node [right] {} (6f);

    \addvmargin{2mm}
\end{tikzpicture} \\
         & & \begin{tikzpicture}[->,>=stealth',shorten >=1pt,auto,node distance=2cm,
                    thick,main node/.style={circle,draw,font=\sffamily\Large\bfseries},sp node/.style={shape=semicircle,draw, shape border rotate=180},so node/.style={shape=semicircle,draw},empty node/.style={font=\sffamily\Large\bfseries},square node/.style={rectangle,draw}]

  \node[main node,scale=0.8] (1) {$N+1$};
  \node[square node,inner sep=0.2cm] (1f) [above of = 1, above=-0.8cm] {$2$ \textbf{fund}};
  \node[sp node,inner sep=0.2cm] (2) [right of=1] {$2 N$};
  \node[so node] (7) [right of=2] {$2 N+2$};
  \node [empty node] (8) [right of=7, left=0cm] {\dots};
  \node [so node] (9) [right of=8, left=-0.35cm] {$2 N+2$};
  \node[sp node,inner sep=0.2cm] (4) [right of=9]{$2N$};
  \node[main node,scale=0.8] (5) [right of=4,right=-0.25cm]{$N+1$};
  \node[square node,inner sep=0.2cm] (5f) [above of = 5, above=-0.8cm] {$2$ \textbf{fund}};
  
  \path[every node/.style={font=\sffamily\small}]
    (1) edge [bend left=20] node [left] {} (2)
    (2) edge[-] node [right] {} (7)
        edge [bend left=20] node [left] {} (1)
    (7) edge[-] node [right] {} (8)
    (8) edge [-] node [right] {} (9)
    (9) edge [-] node [right] {} (4)
    (4) edge [bend left=20] node [right] {} (5)
    (5) edge [bend left=20] node [right] {} (4)
    (1) edge [-] node [right] {} (1f)
    (5) edge [-] node [right] {} (5f);
\end{tikzpicture} \\
         & & \begin{tikzpicture}[->,>=stealth',shorten >=1pt,auto,node distance=2cm,
                    thick,main node/.style={circle,draw,font=\sffamily\Large\bfseries},square node/.style={rectangle,draw},sp node/.style={shape=semicircle,draw, shape border rotate=180},empty node/.style={font=\sffamily\Large\bfseries},square node/.style={rectangle,draw}]

  \node[main node] (1) {$N$};
  \node[main node] (2) [below right of=1] {$2N$};
  \node[main node] (3) [below left of=2] {$N$};
  \node[empty node] (4) [right of=2, left=0cm] {\dots};
  \node[main node] (5) [right of=4, left=0cm] {$2N$};
  \node[sp node, inner sep=0.15cm] (6) [right of=5] {$2N$};
  \node[square node,inner sep=0.2cm] (6f) [above of = 6, above=-1cm] {$2$ \textbf{fund}};

  \path[every node/.style={font=\sffamily\small}]
    (1) edge [bend left=20] node [left] {} (2)
    (2) edge [bend left=20] node [left] {} (1)
        edge [bend left=20] node [left] {} (3)
        edge [bend left=20] node [left] {} (4)
    (3) edge [bend left=20] node [right] {} (2)
    (4) edge [bend left=20] node [right] {} (5)
        edge [bend left=20] node [right] {} (2)
    (5) edge [bend left=20] node [right] {} (6)
        edge [bend left=20] node [right] {} (4)
    (6) edge [bend left=20] node [right] {} (5)
        edge [-] node [right] {} (6f);
\end{tikzpicture} \\
         \cline{2-3}
         & \multirow[b]{2}{3cm}{\centering{Odd}} & \begin{tikzpicture}[baseline=0,->,>=stealth',shorten >=1pt,auto,node distance=1.5cm,
                    thick,main node/.style={circle,draw,font=\sffamily\Large\bfseries},sp node/.style={shape=semicircle,draw, shape border rotate=180},so node/.style={shape=semicircle,draw},empty node/.style={font=\sffamily\Large\bfseries},square node/.style={rectangle,draw}]

  \node[so node,scale = 0.8] (2) [] {$2N+2$};
  \node[sp node] (1) [above left of=2] {$N$};
  \node[square node,inner sep=0.2cm] (1f) [left of = 1] {$1$ \textbf{fund}};
  \node[sp node] (3) [below left of=2] {$N$};
  \node[square node,inner sep=0.2cm] (3f) [left of = 3] {$1$ \textbf{fund}};
  \node[sp node] (7) [right of=2] {$2N$};
  \node[empty node] (8) [right of = 7] {\dots};
  \node[sp node] (9) [right of=8] {$2N$};
  \node[main node,scale=0.7] (5) [right of=9, right=0cm] {$N+1$};
  \node[square node, inner sep=0.2cm] (5f) [above of=5,above=-0.7cm] {$2$ \textbf{fund}};

  \path[every node/.style={font=\sffamily\small}]
    (1) edge[-] node [left] {} (2)
    (2) edge[-] node [right] {} (7)
    (3) edge[-] node [left] {} (2)
    (7) edge[-] node [right] {} (8)
    (8) edge[-] node [right] {} (9)
    (1) edge[-] node [right] {} (1f)
    (3) edge[-] node [right] {} (3f)
    (9) edge[bend left=20] node [right] {} (5)
    (5) edge[bend left=20] node [right] {} (9)
    (5) edge[-] node [right] {} (5f);

    \addvmargin{2mm}
\end{tikzpicture}\\
         & & \begin{tikzpicture}[->,>=stealth',shorten >=1pt,auto,node distance=2cm,
                    thick,main node/.style={circle,draw,font=\sffamily\Large\bfseries},square node/.style={rectangle,draw},sp node/.style={shape=semicircle,draw, shape border rotate=180},empty node/.style={font=\sffamily\Large\bfseries},square node/.style={rectangle,draw}]

  \node[main node] (1) {$N$};
  \node[main node] (2) [below right of=1] {$2N$};
  \node[main node] (3) [below left of=2] {$N$};
  \node[empty node] (4) [right of=2, left=0cm] {\dots};
  \node[main node] (5) [right of=4, left=0cm] {$2N$};
  \node[square node, inner sep=0.2cm] (5e) [right of=5] {$1$ \textbf{Asymm}};
  \node[square node,inner sep=0.2cm] (5f) [above of = 5, above=-1cm] {$2$ \textbf{fund}};

  \path[every node/.style={font=\sffamily\small}]
    (1) edge [bend left=20] node [left] {} (2)
    (2) edge [bend left=20] node [left] {} (1)
        edge [bend left=20] node [left] {} (3)
        edge [bend left=20] node [left] {} (4)
    (3) edge [bend left=20] node [right] {} (2)
    (4) edge [bend left=20] node [right] {} (5)
        edge [bend left=20] node [right] {} (2)
    (5) edge [bend left=20] node [right] {} (4)
        edge [-] node [right] {} (5e)
        edge [-] node [right] {} (5f);
\end{tikzpicture}\\
         \hline
    \end{tabular}
\caption{Summary of orientifolds that contain an O7-plane, and their corresponding quivers. These are the orientifolds that require the introduction of D7-branes. Here we set the size of each gauge group factor to the appropriate value to ensure conformal symmetry. Note that we can move the hypermultiplets transforming in the \textbf{fundamental} representation of a gauge group factor to a different node of the quiver. Doing so changes the relative sizes of the nodes.}
\label{O7Table}
\end{table}

\begin{table}
\centering
    \begin{tabular}{|| >{\centering\Large} m{4cm} | >{\centering\arraybackslash} m{10cm} ||}
         \hline
         Orientifold Group & \Large{Quiver Diagram(s)} \\
         \hline\hline
         Tetra $\subset$ Octa & \begin{tikzpicture}[baseline=0,->,>=stealth',shorten >=1pt,auto,node distance=2cm,
                    thick,main node/.style={circle,draw,font=\sffamily\Large\bfseries},square node/.style={rectangle,draw},sp node/.style={shape=semicircle,draw, shape border rotate=180},so node/.style={shape=semicircle,draw},
                    empty node/.style={font=\sffamily\Large\bfseries}]

  \node[sp node, inner sep = 0.2cm] (1) {$6N$};
  \node[so node, inner sep=0.05cm] (2a) [above of=1, above = -1.4cm] {$4N + \frac{4}{3}$};
  \node[so node, inner sep=0.05cm] (2b) [below left of=1, above = 0.25 cm] {$4N + \frac{4}{3}$};
  \node[so node, inner sep=0.05cm] (2c) [below right of=1, above = 0.25 cm] {$4N + \frac{4}{3}$};
  \node[sp node, inner sep=0.05cm] (3a) [above of=2a, above = -1.05cm] {$2N - \frac{4}{3}$};
  \node[sp node, inner sep=0.05cm] (3b) [below left of=2b, above = 0.17 cm] {$2N - \frac{4}{3}$};
  \node[sp node, inner sep=0.05cm] (3c) [below right of=2c, above = 0.17 cm] {$2N - \frac{4}{3}$};

  \path[every node/.style={font=\sffamily\small}]
    (1) edge [-] node [left] {} (2a)
        edge [-] node [left] {} (2b)
        edge [-] node [left] {} (2c)
    (2a)edge [-] node [left] {} (3a) 
    (2b)edge [-] node [left] {} (3b)
    (2c)edge [-] node [left] {} (3c);

    \addvmargin{2mm}
\end{tikzpicture}
         \hspace{1cm}
         \begin{tikzpicture}[baseline=0,->,>=stealth',shorten >=1pt,auto,node distance=2cm,
                    thick,main node/.style={circle,draw,font=\sffamily\Large\bfseries},square node/.style={rectangle,draw},sp node/.style={shape=semicircle,draw, shape border rotate=180},so node/.style={shape=semicircle,draw},
                    empty node/.style={font=\sffamily\Large\bfseries}]

  \node[so node, inner sep = 0.2cm] (1) {$6N$};
  \node[sp node, inner sep=0.05cm] (2a) [above of=1, above = -1.2cm] {$4N - \frac{4}{3}$};
  \node[sp node, inner sep=0.05cm] (2b) [below left of=1, above = 0.05 cm] {$4N - \frac{4}{3}$};
  \node[sp node, inner sep=0.05cm] (2c) [below right of=1, above = 0.05 cm] {$4N - \frac{4}{3}$};
  \node[so node, inner sep=0.05cm] (3a) [above of=2a, above = -1.3cm] {$2N + \frac{4}{3}$};
  \node[so node, inner sep=0.05cm] (3b) [below left of=2b, above = 0.3 cm] {$2N + \frac{4}{3}$};
  \node[so node, inner sep=0.05cm] (3c) [below right of=2c, above = 0.3 cm] {$2N + \frac{4}{3}$};

  \path[every node/.style={font=\sffamily\small}]
    (1) edge [-] node [left] {} (2a)
        edge [-] node [left] {} (2b)
        edge [-] node [left] {} (2c)
    (2a)edge [-] node [left] {} (3a) 
    (2b)edge [-] node [left] {} (3b)
    (2c)edge [-] node [left] {} (3c);

    \addvmargin{2mm}
\end{tikzpicture} \\
         \hline
         $\mathbb{Z}_2^{\Omega'} \times \text{Tetra}$ & \begin{tikzpicture}[baseline=0,->,>=stealth',shorten >=1pt,auto,node distance=2.0cm,
                    thick,main node/.style={circle,draw,font=\sffamily\Large\bfseries},square node/.style={rectangle,draw},sp node/.style={shape=semicircle,draw, shape border rotate=180},so node/.style={shape=semicircle,draw},
                    empty node/.style={font=\sffamily\Large\bfseries}]

  \node[sp node, scale=1.3] (1) {$3N$};
  \node[main node] (2) [right of=1] {$2N$};
  \node[main node] (3) [right of=2] {$N$};
  \node[so node] (5) [left of=1] {$2N+4$};
  \node[sp node] (6) [left of=5] {$N+4$};
  \node[square node, inner sep=0.2cm](6f)[left of=6]{$4$ \textbf{fund}};

  \path[every node/.style={font=\sffamily\small}]
    (1) edge [bend left=20] node [left] {} (2)
        edge [-] node [left] {} (5)
    (2) edge [bend left=20] node [right] {} (3)
    (2) edge [bend left=20] node [left] {} (1)
    (3) edge [bend left=20] node [right] {} (2)
    (5) edge [-] node [left] {} (6)
    (6) edge [-]  node [left] {} (6f);

    \addvmargin{2mm}
\end{tikzpicture} \\
         \hline
         \multirow[b]{2}{4cm}{\centering{$\mathbb{Z}_2^{\Omega'} \times \text{Octa}$}} & \begin{tikzpicture}[baseline=0,->,>=stealth',shorten >=1pt,auto,node distance=2.0cm,
                    thick,main node/.style={circle,draw,font=\sffamily\Large\bfseries},square node/.style={rectangle,draw},sp node/.style={shape=semicircle,draw, shape border rotate=180},so node/.style={shape=semicircle,draw},
                    empty node/.style={font=\sffamily\Large\bfseries}]

  \node[sp node, scale=1.3] (1) {$4N$};
  \node[main node] (2) [right of=1] {$3N$};
  \node[main node] (3) [right of=2] {$2N$};
  \node[main node] (4) [right of=3] {$N$};
  \node[so node] (5) [left of=1] {$2N+4$};
  \node[square node, inner sep=0.2cm](5f)[left of=5]{$2$ \textbf{fund}};

  \path[every node/.style={font=\sffamily\small}]
    (1) edge [bend left=20] node [left] {} (2)
        edge [-] node [left] {} (5)
    (2) edge [bend left=20] node [right] {} (3)
    (2) edge [bend left=20] node [left] {} (1)
    (3) edge [bend left=20] node [right] {} (2)
    (3) edge [bend left=20] node [right] {} (4)
    (4) edge [bend left=20] node [right] {} (3)
    (5) edge [-]  node [left] {} (5f);

    \addvmargin{2mm}
\end{tikzpicture} \\
         & \begin{tikzpicture}[->,>=stealth',shorten >=1pt,auto,node distance=2cm,
                    thick,main node/.style={circle,draw,font=\sffamily\Large\bfseries},square node/.style={rectangle,draw},sp node/.style={shape=semicircle,draw, shape border rotate=180},so node/.style={shape=semicircle,draw},
                    empty node/.style={font=\sffamily\Large\bfseries}]

  \node[so node, inner sep=0.2cm] (1) {$4N$};
  \node[sp node] (2a) [above of=1, above = -1cm] {$2N-2$};
  \node[sp node] (2b) [left of=1] {$3N-1$};
  \node[sp node] (2c) [right of=1] {$3N-1$};
  \node[so node] (3b) [left of=2b] {$2N+2$};
  \node[so node] (3c) [right of=2c] {$2N+2$};
  \node[sp node] (4b) [left of=3b] {$N+1$};
  \node[sp node] (4c) [right of=3c] {$N+1$};
  \node[square node,inner sep=0.2cm](5b) [above of=4b, above=-1.2cm] {$2$ \textbf{fund}};
  \node[square node,inner sep=0.2cm](5c) [above of=4c, above=-1.2cm] {$2$ \textbf{fund}};

  \path[every node/.style={font=\sffamily\small}]
    (1) edge [-] node [left] {} (2a)
        edge [-] node [left] {} (2b)
        edge [-] node [left] {} (2c)
    (2b) edge [-] node [left] {} (3b)
    (2c) edge [-] node [left] {} (3c)
    (3b) edge [-] node [left] {} (4b)
    (3c) edge [-] node [left] {} (4c)
    (4b) edge [-] node [right] {} (5b)
    (4c) edge [-] node [right] {} (5c);
\end{tikzpicture}\\
         \hline
         $\mathbb{Z}_2^{\Omega'} \times \text{Icosa}$ & \begin{tikzpicture}[baseline=0,->,>=stealth',shorten >=1pt,auto,node distance=2cm,
                    thick,main node/.style={circle,draw,font=\sffamily\Large\bfseries},square node/.style={rectangle,draw},sp node/.style={shape=semicircle,draw, shape border rotate=180},so node/.style={shape=semicircle,draw},
                    empty node/.style={font=\sffamily\Large\bfseries}]

  \node[so node, inner sep=0.2cm] (1) {$6N$};
  \node[sp node] (2a) [above of=1, above = -1cm] {$3N - 2$};
  \node[sp node] (2b) [left of=1] {$4N - \frac{4}{3}$};
  \node[sp node] (2c) [right of=1] {$5N - \frac{2}{3}$};
  \node[so node] (3b) [left of=2b] {$2N + \frac{4}{3}$};
  \node[so node] (3c) [right of=2c] {$4N + \frac{8}{3}$};
  \node[sp node] (4c) [right of=3c] {$3N + 2$};
  \node[so node](5c) [right of=4c] {$2N + \frac{16}{3}$};
  \node[sp node](6c) [right of=5c] {$N + \frac{14}{3}$};
  \node[square node, inner sep=0.2cm](7c) [above of=6c, above=-1.2cm]{$4$ \textbf{fund}};

  \path[every node/.style={font=\sffamily\small}]
    (1) edge [-] node [left] {} (2a)
        edge [-] node [left] {} (2b)
        edge [-] node [left] {} (2c)
    (2b) edge [-] node [left] {} (3b)
    (2c) edge [-] node [left] {} (3c)
    (3c) edge [-] node [left] {} (4c)
    (4c) edge [-] node [right] {} (5c)
    (5c) edge [-] node [right] {} (6c)
    (6c) edge [-] node [right] {} (7c);

    \addvmargin{2mm}
\end{tikzpicture}\\
         \hline
    \end{tabular}
\caption{Summary of orientifolds that involve the exceptional subgroups of $\SU(2)$, and their corresponding quivers. Here we set the size of each gauge group factor to the appropriate value to ensure conformal symmetry. Note that we can move the hypermultiplets transforming in the \textbf{fundamental} representation of a gauge group factor to a different node of the quiver. Doing so changes the relative sizes of the nodes.}
\label{ExceptionalTable}
\end{table}

\clearpage

\bibliographystyle{JHEP}
\bibliography{main}

\end{document}